\documentclass[english]{article}
\usepackage[T1]{fontenc}
\usepackage[latin9]{inputenc}
\usepackage{color}
\usepackage{textcomp}
\usepackage{url}
\usepackage{amsmath}
\usepackage{amssymb}
\usepackage{graphicx}
\usepackage{esint}

\makeatletter

\usepackage{fullpage}
\usepackage{tocbibind}

\makeatother

\usepackage{babel}
\begin{document}
\title{Computation of extremes values of time averaged observables in climate
models with large deviation techniques }
\author{Francesco Ragone\textsuperscript{1}, Freddy Bouchet\textsuperscript{1}}

\maketitle
\textsuperscript{1}Univ Lyon, Ens de Lyon, Univ Claude Bernard, CNRS,
Laboratoire de Physique, Lyon, France
\begin{abstract}
One of the goals of climate science is to characterize the statistics
of extreme and potentially dangerous events in the present and future
climate. Extreme events like heat waves, droughts, or floods due to
persisting rains are characterized by large anomalies of the time
average of an observable over a long time. The framework of Donsker\textendash Varadhan
large deviation theory could therefore be useful for their analysis.
In this paper we discuss how concepts and numerical algorithms developed
in relation with large deviation theory can be applied to study extreme,
rare fluctuations of time averages of surface temperatures at regional
scale with comprehensive numerical climate models. We study the convergence
of large deviation functions for the time averaged European surface
temperature obtained with direct numerical simulation of the climate
model Plasim, and discuss their climate implications. We show how
using a rare event algorithm can improve the efficiency of the computation
of the large deviation rate functions. We discuss the relevance of
the large deviation asymptotics for applications, and we
show how rare event algorithms can be used also to improve the statistics
of events on time scales shorter than the one needed for reaching
the large deviation asymptotics.

\end{abstract}

\section{Introduction}

The study of high impact rare events, like extreme droughts, heat
waves, rainfalls and storms, is a major topic of interest in climate
science. Extreme events can have a severe impact on ecosystems and
socio-economic systems \cite{aghakouchak_extremes_2012,IPCC_managing_2012,IPCC_climate_2013},
and it is crucial to better understand their dynamics and statistical
properties. A general problem is that it is difficult to sample a
sufficient amount of rare extreme events to have a robust statistics.
Observational records are typically too short to study events with
return times longer than a few decades. State of the art general circulation
models are computationally extremely expensive, and can be run at
most for a few thousands of years, making it unrealistic to study
even approximately extreme events with return times longer than a
century.

In climate science several techniques are adopted to compensate for
the lack of data. The general idea is to extract informations about
rare, unobserved events from the limited but available statistics
of less rare, observed events. Purely statistical approaches are
usually framed in the context of extreme value theory \cite{Coles2001,Ghil&al2011,Lucarini&al2016book}.
Stochastic weather generators provide to some extent a hybrid statistical-dynamical
approach \cite{Wilks1999,Ailliot&al2015}. A new approach entirely
based on the dynamics of numerical models was proposed in \cite{Ragone&al2018},
where we have introduced the use of rare event algorithms to improve
the sampling efficiency of climate models. These techniques allow
to increase the number of extreme events observed for a given computational
cost, by generating trajectories that are real solutions of the equations
of the model, without additional statistical assumptions.

Different types of extreme events have different spatio-temporal characteristics.
Events like wind storms or flash floods are transient, typically localized
phenomena due to large fluctuations of an observable on short time
scales (from a few hours to a few days) compared to the typical time
scale of the synoptic variability. Events like heat waves or floods
due to persisting rains are larger spatial scale phenomena characterized
by large anomalies of the time average of an observable over longer
time scales (from several weeks to months).

A statistical framework to analyze time persistent events is provided
by Donsker-Varadhan large deviation theory. Large deviation theory
deals in general with the exponential decay of probabilities of fluctuations
in random systems, providing an extension of the law of large numbers
and central limit theorem \cite{Touchette2009}. Typically one obtains
a large deviation scaling as asymptotic behavior of probability distributions
depending on a small parameter. Donsker-Varadhan large deviation theory
is a particular case of large deviation theory, which deals with the
scaling of the statistics of time averages over a time $T$. It predicts
the asymptotic behavior of the probability distribution function of
the time average of an observable for any large enough $T$ (where
the small parameter is given by $1/T$), described by a large deviation
rate function. Establishing a large deviation result for a climatic
observable would mean that the probability of extremely rare fluctuations
of the time average of that observable can be inferred from the probability
of much less rare events. For example, the probability of having a
(very rare) heat wave characterized by a value $a$ of the temperature
anomaly over several months could be obtained from the probability
of having a (much less rare) heat wave characterized by the same value
$a$ of the temperature anomaly over just a few weeks.

With the exception of a few works on multifractal modeling of rainfall
\cite{Veneziano&al2009}, the use of large deviation theory
to study climate extremes has not been considered until very recently.
In \cite{Ragone&al2018} we used a rare event algorithm developed
for the computation of Donsker-Varadhan type large deviation functions,
and applied it to study rare heat waves, although for durations shorter
than what necessary to be in the large deviation limit. \cite{Galfi&al2019}
recently performed a comparison of extreme value theory and large
deviation theory based approaches to study time and space averages
of climatic observables in an idealized general circulation model.
Considering the increasing interest in time persistent climate extremes, it is of interest to explore
more in depth the applicability of large deviation theory to study
rare fluctuations of time averages of climatic observables, and to
discuss the methodological challenges one faces to perform this type
of analysis.

The first step of an empirical analysis of the large deviations of
the time average of an observable is to determine the minimum length
of the averaging period $T$ for which the convergence to the large
deviation limit is satisfied up to an acceptable degree of accuracy.
The second step is to determine if the data available are enough to
study the non Gaussian tails of the rate function for the chosen range
of values of $T$. \cite{Rohwer&al2015} have recently provided a
systematic analysis of the convergence of statistical estimators of
large deviation functions of sums of independent and identically distributed
random variables, including a detailed study of the maximum range
of fluctuations for which the rate functions can be computed, given
a finite sample of data. These results in principle could be used,
with some additional considerations, to analyse the large fluctuations of the time averages
of an observable from time series of a dynamical process \cite{Rohwer&al2015},
and could therefore be of interest to perform precise large deviation
analysis of climatic observables. 

Given the typical limitations in
the amount of available data in these applications, it is likely that
it will be rather difficult to go substantially (if at all) beyond
the Gaussian regime given by the central limit theorem. In this case,
one could use rare event algorithms dedicated to compute Donsker-Varadhan
type large deviation functions in numerical models that have been
developed in recent years \cite{giardina_direct_2006,lecomte_numerical_2007},
and very recently applied to study heat waves in a climate model \cite{Ragone&al2018}.

The tools to perform a large deviation analysis of time averages of
climatic observables are thus available; however, a systematic description
and evaluation of such tools specifically framed for climate studies
is still lacking. In this paper we describe how to use large deviation principles and
rare event algorithms in a climatic study. The paper is structured
as follows. First we introduce the basic formalism of Donsker-Varadhan
large deviation theory. Then we provide a detailed description of
how to compute large deviation functions from time series of an observable
of a dynamical system, following \cite{Rohwer&al2015}. We give a
practical example of such analysis, computing large deviation functions
of the time average of the European surface temperature from long
simulations with the climate model Plasim \cite{Fraedrich&al2005}.
We then show how the tails of the large deviation functions can be
computed very efficiently using the rare event algorithm we have used
in \cite{Ragone&al2018}, giving here more details about the method and
focusing on the role of large deviation theory. Finally we discuss
the potential for further studies.

\section{Large deviation theory for time averaged observables\label{subsec:Large-deviations-for}}

In this section we introduce large deviation theory for time averaged
observables and the related notation. We consider $X(t)$ the time
dependent state of the dynamics of a climate model, which is a deterministic
dynamical system. In general $X\in\mathbb{R}^{n}$ and $\left\{ X(t)\right\} _{t\geq0}$
is a Markov process. We consider a generic observable $A:\mathbb{R}^{n}\rightarrow\mathbb{R}$,
of which we want to study the statistics. For example, in this paper
we will take $A(X(t))$ as the time dependent surface temperature
averaged over Europe.

For ergodic systems, the time average $a=\frac{1}{T}\int_{0}^{T}A(X(t))\,\mbox{d}t$
converges in the limit of large $T$ to the ergodic average $\mathbb{E}(A)$.
Moreover, when some mixing hypothesis are verified, the central limit
theorem guarantees that, for large $T$, typical fluctuations of $a$
are of order $\sqrt{T}$ and are Gaussian. More precisely, the probability
density function of $\left[a-\mathbb{E}(A)\right]/\sqrt{T}$ is a
Gaussian distribution function with mean $\mathbb{\mu=E}(A)$ and
variance $\int_{0}^{\infty}\mathbb{E}\left[(A(X(t))-\mu)(A(X(0))-\mu)\right]\text{d}t=\sigma^{2}\tau_{c}$
where $\sigma^{2}=\mathbb{E}\left[(A(X(0))-\mu)(A(X(0))-\mu)\right]$
is the variance of $A$ and the previous equality
is a definition of the autocorrelation time $\tau_{c}$.

In many cases we are interested in events much rarer than those described
by a Gaussian approximation. It is then useful to consider fluctuations
of $a$ that are of order $T$, rather than $\sqrt{T}$. For these
large fluctuations, a generalization of the central limit theorem,
called a large deviation result \cite{Touchette2009}, states that
\begin{equation}
\rho(a,T)\underset{T\rightarrow\infty}{\asymp}\mbox{e}^{-TI\left[a\right]},\label{eq:Large_Deviation_a}
\end{equation}
where the non negative function $I(a)$ is called large deviation
rate function. The symbol $f\underset{T\rightarrow\infty}{\asymp}g$
stands for logarithmic equivalence, that is $\log \left[f\right]\underset{T\rightarrow\infty}{\sim}\log\left[g\right]$.
Then (\ref{eq:Large_Deviation_a}) is equivalent to 

\begin{equation}
I(a)=\lim_{T\rightarrow\infty}I(a,T),\,\,\,\mbox{with}\,\,\,I(a,T)=-\frac{1}{T}\log\left[\rho(a,T)\right].\label{eq:rate_function}
\end{equation}
Such a large deviation result is valid for mixing enough dynamics
and observables with probability distribution function that decay sufficiently fast for large values of the observable. For a Gaussian process with exponential correlation function, the autocorrelation time and the mixing time are of the same order of magnitude. However, in more complex dynamics the picture can be more complicated.
Sufficient conditions are given by Donsker-Varadhan's
theory for Markov processes \cite{donsker1975asymptotic,dembo2001large} or by \cite{young1990large,kifer1990large} for
dynamical systems. The more general result $\rho(a,T)\underset{T\rightarrow\infty}{\sim}C(a,T\text{)}\mbox{e}^{-TI(a)}$
would imply $\rho(a,T)\underset{T\rightarrow\infty}{\asymp}\mbox{e}^{-TI(a)}$
if $C(a,T)$ increases less than exponentially for large $T$. The
logarithmic equivalence thus means that the prefactor $C$ is subdominant
in the large $T$ limit with respect to the the exponential term,
and is not determined. 

From (\ref{eq:Large_Deviation_a}), we see that the minimum of $I(a)$
is attained at the most probable values. If we assume that there is
a unique most probable $a_{m}$, then $I(a)\geq I(a_{m})=0$ for any
$a$. If $I(a)>0$ for any $a\neq a_{m}$, then $\rho(a,T)$ concentrate
exponentially close to $a_{m}=\mu=\mathbb{E}(A)$. When a large deviation
result holds, the asymptotic behavior of $\rho(a,T)$, that in general
depends on both $a$ and $T$, is summarized by a single function
$I(a)$. The large deviation asymptotics is then a huge simplification
and the probability of the fluctuations of $a$ for very large values
of $T$ can be determined from fluctuations observed for smaller values
of $T$.

Large deviation theory for time averaged observables could be relevant
in all those cases in which an extreme event is characterized by its
anomalous persistence in time (e.g. heat waves and cold spells, windy
seasons, droughts, accumulation of rainfall in flood prone regions,
etc.). In particular, if the considered observable is a flux with
respect to time (e.g. precipitation, greenhouse gases emissions),
then the accumulated anomaly is the actual surplus of the quantity
accumulated during the observation period.

Rather than dealing directly with the probability distribution function
$\rho(a,T)$, it is often useful and practical to compute the scaled
cumulant generating function 
\begin{equation}
\lambda(k)=\lim_{T\rightarrow\infty}\lambda(k,T),\,\,\,\mbox{with}\,\,\,\lambda(k,T)=\frac{1}{T}\log\left[\mathbb{E}\left(\mbox{e}^{k\int_{0}^{T}A(X(t))\,\mbox{d}t}\right)\right].\label{eq:Scaled_cumulant_generating_function}
\end{equation}
We can notice that $\int\textrm{d}a\,e^{T[ka-I(a,T)]}\underset{T\rightarrow\infty}{\asymp}\mbox{e}^{T\lambda(k,T)}$,
where we have used (\ref{eq:Scaled_cumulant_generating_function})
and (\ref{eq:Large_Deviation_a}). Since $T$ is very large, the Laplace
integral on the left hand side is dominated by the supremum of the
argument of the exponential (analogously to a saddle point approximation),
and in the limit of $T$ going to infinity we have
\begin{equation}
\lambda(k)=\sup_{a}\left\{ ka-I(a)\right\} .\label{eq:Legendre_Fenchel}
\end{equation}
Such a relation between $I(a)$ and $\lambda(k)$ is called a Legendre\textendash Fenchel
transform. When $I(a)$ is a convex function and differentiable, or
equivalently when $\lambda(k)$ is differentiable, the Legendre\textendash Fenchel
transform can be inverted and $I(a)=\sup_{k}\left\{ ka-\lambda(k)\right\} $.
The hypothesis under which this heuristic derivation is valid are
provided by the Gärtner\textendash Ellis theorem \cite{ellis2007entropy}. On domains for which the Legendre\textendash Fenchel
is invertible and $\lambda(k)$ differentiable, the variational problem
(\ref{eq:Legendre_Fenchel}) gives $I(a)=k(a)a-\lambda(k(a))$, where
$k(a)$ is given by $a=\lambda'(k(a))$. 

Assuming that $I(a)$ is twice differentiable, we can obtain informations
about the Gaussian fluctuations directly from $I(a)$. Performing
a Taylor expansion of (\ref{eq:Large_Deviation_a}), one obtains that
$\frac{1}{\sqrt{T}}\int_{0}^{T}\left[A(X(t))-\mathbb{E}(A)\right]\,\mbox{d}t$
is asymptotically Gaussian, with variance $\sigma=1/I''(a)$. This
can be seen also expanding the scaled cumulant generating function
in powers of $k$, which gives $\lambda(k)=\tau_{c}\sigma^{2}k^{2}+O(k^{3})$,
where $\tau_{c}$ is the autocorrelation time of the observable defined
above. Consequently, $\lambda^{'}(k)=2\tau_{c}\sigma^{2}k+O(k^{2})$
and $I(a)=a^{2}/4\tau_{c}\sigma^{2}+O(a^{3})$. However, $I(a)$ contains
more information than just the average and the Gaussian fluctuations:
the next derivatives of $I(a)$ are related to higher order cumulants,
and $I(a)$ for large values of $a$ characterizes rare events beyond
the Gaussian approximation. Clearly, the main interest in analyzing
large deviation functions lies in having access to the non Gaussian
parts of their tails.

\section{Estimate of large deviation functions with direct sampling\label{subsec:Estimate-of-large}}

\subsection{Direct estimate of large deviations from time series\label{subsec:Direct-estimate-of-large-deviations}}

In this section we describe how to compute empirically large deviation
functions from timeseries of $A(X(t))$, following closely \cite{Rohwer&al2015}. The
presentation is kept as simple as possible, and aims at providing
a clear recipe that can be reproduced for any application with complex
numerical models. More technical details about the convergence of
the estimators are presented in Appendix \ref{sec:Direct-estimate-Appendix}.

Computing estimates of large deviation functions from the time correlated
output of a complex dynamical system is not trivial. In general the
applicability of the large deviation scaling depends on whether the
time scales that characterize the persistence of the rare events of
interest are large enough such that they belong to the asymptotic
regime. Heuristically, a minimal requirement is that $T$ is much
larger than the autocorrelation time $\tau_{c}$ of the time series
of the observable. However the full answer to this question strongly
depends on the structure of the correlations of $A(X(t))$, and on
the overall distribution of the process. When performing an empirical
estimation, the convergence of (\ref{eq:rate_function}) and (\ref{eq:Scaled_cumulant_generating_function})
has to be analysed case by case. 

Let us suppose that we have a long simulation obtained running a climate
model for a time $T$, and that we want to perform a large deviation
analysis of an observable $A(X(t))$. We divide the time series in
$N_{b}=T/\tau_{b}$ blocks of length $\tau_{b}$, and we consider
the time average of the chosen observable $A(X(t))$ in the $j$-th
block

\begin{equation}
A_{\tau_{b}}^{j}=\frac{1}{\tau_{b}}\intop_{(j-1)\tau_{b}}^{j\tau_{b}}A(X(t))\,\text{d}t
\end{equation}

Under mixing conditions for the process $A(X(t))$, for a sufficiently
large $\tau_{b}$ the random variable $A_{\tau_{b}}$ can be considered
by block averaging as a sum of independent and identically distributed
(iid) random variables, for which explicit results for the convergence
of estimators of large deviation functions have been obtained \cite{Rohwer&al2015}. 
Euristically one expects that the integral in each block corresponds to a sum over $\tau_b/\tau_c$ values.
The $N_b$ values from the original time series are then taken as independent realizations of such sum of iid random variables.
This approach allows  to study precisely the convergence of the
estimates of the large deviation functions. Note that the same approach
can be followed if instead of a single long simulation we have an
ensemble of shorter simulations each of length $\tau_b$.

In a climate application we are interested in computing the rate function
$I(a)$. However, the direct estimation of the rate function is not
usually the best way to proceed \cite{Rohwer&al2015}, as it is difficult
to provide quantitative arguments to justify the convergence of estimators
of probability density functions. A more precise way of proceeding
is to compute the scaled cumulant generating function first, and then
to compute the rate function as its Legendre\textendash Fenchel transform.
Following this approach, the first step is to compute an estimate
of the generating function approximating the expectation value with
the average over the $N_{b}$ blocks in the limit of large $N_{b}$

\begin{equation}
\hat{G}(k,\tau_{b},N_{b})=\frac{1}{N_{b}}\sum_{j=1}^{N_{b}}e^{kA_{\tau_{b}}^{j}}.\label{eq:GF_empirical}
\end{equation}
 An estimate of the scaled cumulant generating function can then be
computed as

\begin{equation}
\hat{\lambda}(k,\tau_{b},N_{b})=\frac{1}{\tau_{b}}\log\hat{G}(k,\tau_{b},N_{b})\label{eq:SCGF_empirical}
\end{equation}
Given a value of $k$ and estimate of the derivative of the scaled
cumulant generating function $\lambda'(k)=a(k)$ is computed as

\begin{equation}
\hat{a}(k,\tau_{b},N_{b})=\frac{\sum_{j=1}^{N_{b}}A_{\tau_{b}}^{j}e^{kA_{\tau_{b}}^{j}}}{\sum_{j=1}^{N_{b}}e^{kA_{\tau_{b}}^{j}}},\label{eq:rate_function_argument_empirical_estimator_parametric}
\end{equation}
and eventually the estimate of the rate function is computed as

\begin{equation}
\hat{I}(\hat{a}(k,\tau_{b},N_{b}),\tau_{b},N_{b})=k\hat{a}(k,\tau_{b},N_{b})-\hat{\lambda}(k,\tau_{b},N_{b}).\label{eq:rate_function_empirical_estimator_parametric}
\end{equation}

Note that here we have two limits. In the limit $N_{b}\rightarrow+\infty$,
the estimate $\hat{\lambda}(k,\tau_{b},N_{b})$ converges to $\lambda(k,\tau_{b})=\frac{1}{\tau_{b}}\log\mathbb{E}\left[e^{k\intop_{0}^{\tau_{b}}A(X(t))\textrm{d}t}\right],$
which, in the limit $\tau_{b}\rightarrow+\infty$, converges to the
scaled cumulant generating function $\lambda(k)$ (and consequently
the same holds for the convergence of $\hat{I}(k,\tau_{b},N_{b})$
to $I(a)$). The convergence of both limits has to be checked to ensure
the correct computation of the large deviation functions. In a practical
application one is constrained by the fixed length $T=N_{b}\tau_{b}$
of the time-series, so that one faces a trade off in the choice of
$\tau_{b}$ and $N_{b}$. The appeal of this method over attempting
at a direct estimate of the rate function is that one can check precisely
the convergence, exploiting results on the convergence of estimators
of expectation values of exponentials of sums of random variables
\cite{Rohwer&al2015}. The convergence is limited by two problems.

The first issue is that, due to the finite size of the sample, as
the value of $k$ increases the sum over the realizations is rapidly
dominated by the largest value in the sample. This leads to the artificial
linearization of the tails of the estimate of the scaled cumulant
generating function. Given $\tau_{b}$ and $N_{b}$, one can estimate
upper and lower bounds $k_{c}^{-}(\tau_{b},N_{b})$ and $k_{c}^{+}(\tau_{b},N_{b})$
such that the estimate of generating function converges for $k_{c}^{-}(\tau_{b},N_{b})<k<k_{c}^{+}(\tau_{b},N_{b})$.
There are different possible ways to determine values for $k_{c}^{-}(\tau_{b},N_{b})$
and $k_{c}^{+}(\tau_{b},N_{b})$ given a sample of data, and to estimate
their scaling with $N_{b}$, as discussed in \cite{Rohwer&al2015}.
In this paper we have taken a completely empirical approach, and determined
the bounds by requiring that the relative contribution of the largest
value in the sample to the estimate of the generating function does
not overcome an arbitrary threshold of 50\% (see Appendix \ref{sec:Direct-estimate-Appendix}).
Note that if the observable is bounded, that is if it has an upper
or lower limit, the linearization is not an artifact but it is the
correct behavior of the scaled cumulant generating function.

The second issue is the non-uniform convergence of the estimate when
$k$ is increased, which limits the regions in which statistical errors
can be defined. In order to define a statistical error, one normally
assumes that the distribution of the sum over the $N_{b}$ values
converges to a Gaussian distribution around its mean. Statistical
errors are then defined based on the standard deviation of the distribution.
For of sum of exponentials of random variables, like in our case,
this is true only on half of the convergence region of the estimator
\cite{Rohwer&al2015}. For $k_{c}^{-}(\tau_{b},N_{b})/2<k<k_{c}^{+}(\tau_{b},N_{b})/2$,
the estimators converge, they are Gaussian-distributed, and statistical
errors can be computed from their empirical variance. For $k_{c}^{-}(\tau_{b},N_{b})<k<k_{c}^{-}(\tau_{b},N_{b})/2$
or $k_{c}^{+}(\tau_{b},N_{b})/2<k<k_{c}^{+}(\tau_{b},N_{b})$, the
estimators converge, but they are not Gaussian-distributed, and their
statistical error cannot be determined from the empirical variance.
The estimates in these regions therefore cannot be properly analyzed.
In the inner convergence region, the statistical error on the estimate
of $G(k)$ is computed as the standard deviation of the sample of
values involved in the sum replacing the expectation value. The statistical
errors on the estimates of $\lambda(k)$ and $I(a)$ can then be computed
by error propagation, as described in Appendix \ref{sec:Direct-estimate-Appendix}.

By studying the empirical convergence of the estimators, one can identify
an optimal value (or range of values) of $\tau_{b}$ and $N_{b}$,
and obtain the corresponding best estimates of the large deviation
functions. One is typically interested in the tails of the large deviation
functions, beyond the Gaussian approximation. Correctly estimating
the tails however requires a large amount of data. How large depends
critically on the characteristics of the process under study. The
longer it needs to converge to the large deviation limit, the larger
the block size $\tau_{b}$ has to be. For a fixed length of the available
record this means a smaller number of blocks $N_{b}$, and thus a
poorer statistics and a narrower convergence domain, possibly confined
to the Gaussian region. With observational records the problem of
limited data availability can not be circumvented. With numerical
models, a possible way out is given by rare event algorithms, as discussed
in Section \ref{sec:Estimate-of-large}.

\subsection{Analysis of large deviations of European surface temperature in a
climate model with direct sampling}

We give here a demonstration of the procedure described in the previous
Section, by computing large deviation functions for the average European
surface temperature in the numerical climate model Plasim \cite{Fraedrich&al2005}.
The model is set at a T42 horizontal resolution and 10 levels vertical
resolution, for a total of $O(10^{5})$ degrees of freedom. The model
features a full suite of physical parameterizations (see the Reference
Manual freely available together with the code at \url{http://www.mi.uni-hamburg.de/plasim})
and creates a fairly realistic climate. In order to simplify the analysis
we remove the daily and seasonal cycles from the system, so that the
evolution equations do not explicitly depend on time. The standard
tested version of Plasim already runs without daily cycle. In order
to remove the seasonal cycle we set the boundary conditions (sea surface
temperature, ice coverage, and radiative forcing at top of the atmosphere)
to their climatological values for the 16th of July, so that the model
runs in perpetual summer conditions.

\begin{figure}
\begin{centering}
\includegraphics[height=0.45\textwidth]{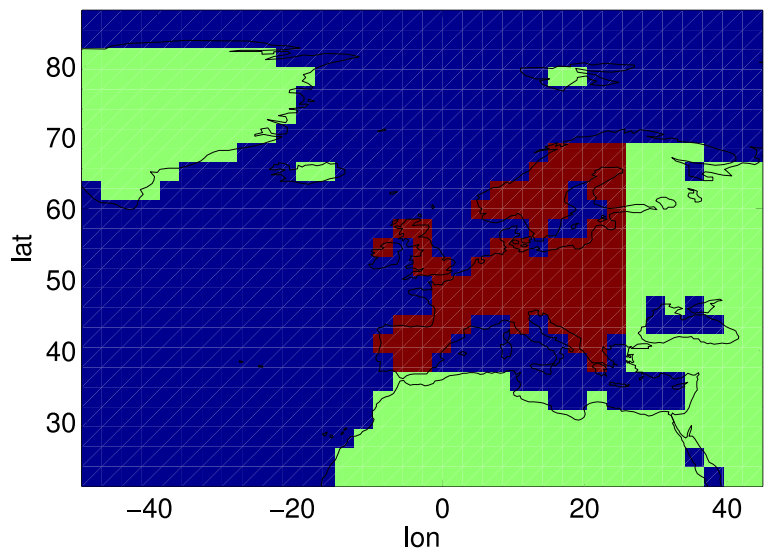}
\end{centering}
\caption{European domain $\Omega$ over which the surface temperature is averaged
(in red). Note that here we show a zoom over a region, but the model
Plasim simulate the dynamics over the whole globe.\label{fig:Map}}
\end{figure}

We consider a 1000 year long simulation. We take as target observable
the European surface temperature $T_{\Omega}(X(t))$, computed as the average of the local surface temperature $T_{s}(\phi,\lambda,t)$ (where $\phi$ is the latitude and $\lambda$ the longitude), over the domain $\Omega$
shown in figure (\ref{fig:Map}), given by the land area included
between 36 °N and 70 °N, and -11 °W and 25 °E. Since the spatial average depends only on time, we explicit in $T_{\Omega}(X(t))$ only the dependence on time of the state of the system, consistently with the notation in the previous section (while in general the state of the system depends on both space and time when we consider an extended system like a climate model).

\begin{figure}
\begin{centering}
a)\includegraphics[height=0.45\textwidth]{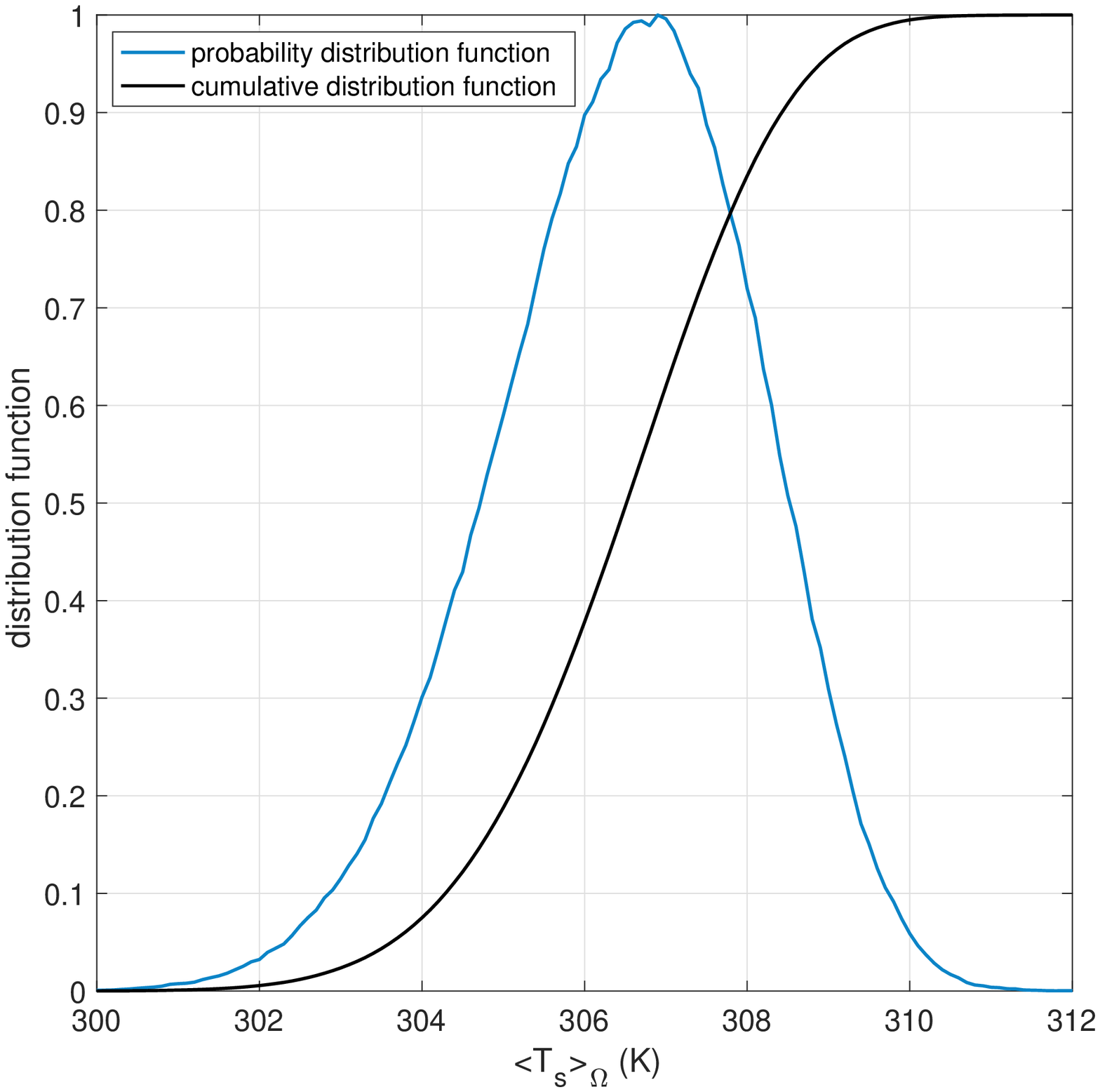}\\
b)\includegraphics[height=0.45\textwidth]{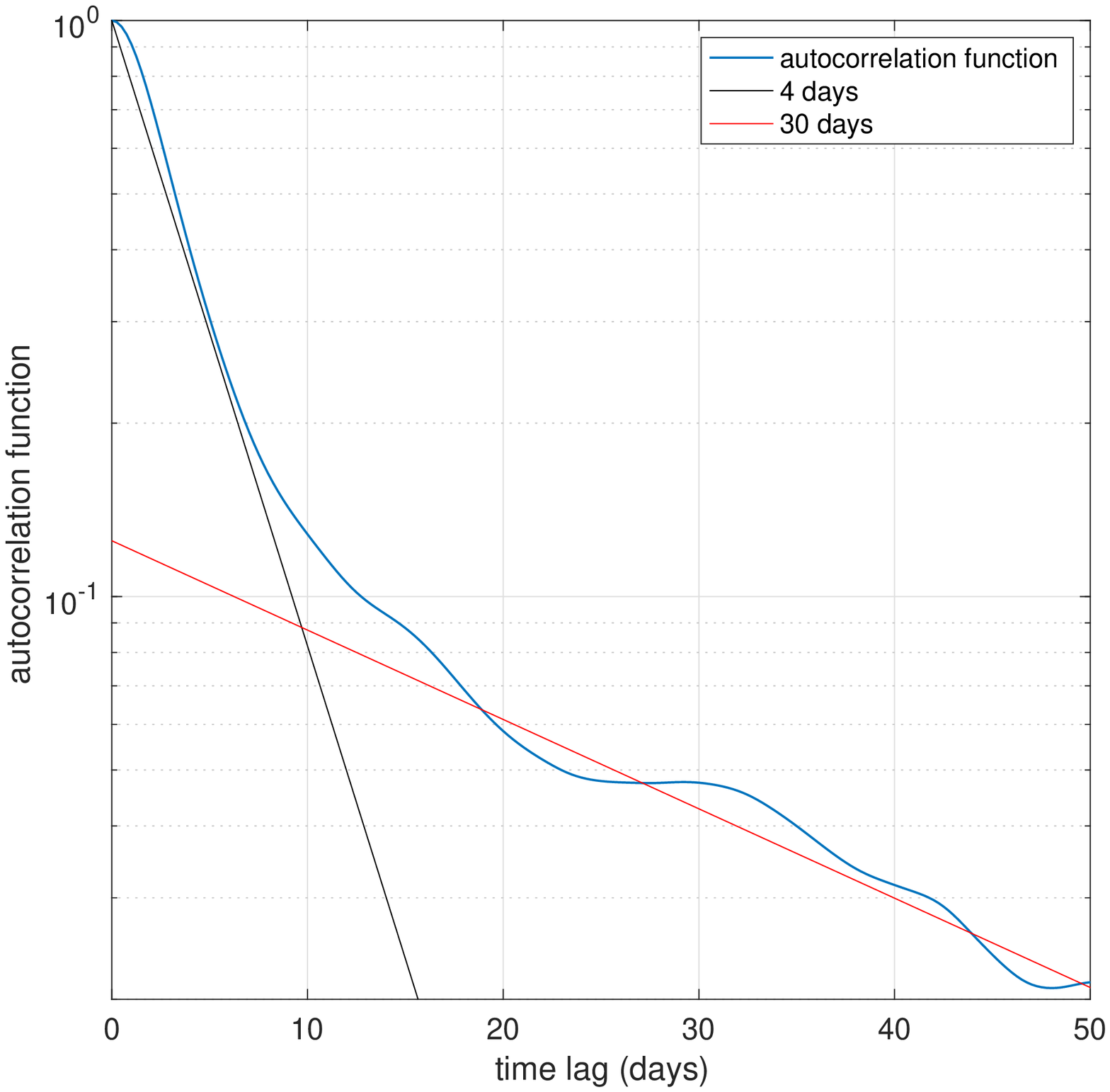}\\
\end{centering}
\caption{a) Probability distribution function (blue) and cumulative distribution
function (black) of the 6 hourly values of the average European surface
temperature $T_{\Omega}(X(t))$. Fro graphical reasons the probability
distribution function has been normalized so that its maximum has
value 1. b) Autocorrelation function of $T_{\Omega}(X(t))$ (blue).
The black and red lines show exponential decays on time scales of
4 and 30 days respectively.\label{fig:a)-Probability-distribution}}
\end{figure}

Figure (\ref{fig:a)-Probability-distribution}a) shows the probability
density function (normalized so that its maximum has value 1) and
cumulative distribution function of the 6 hourly values of $T_{\Omega}(X(t))$.
The spatially averaged temperature has mean $\mu\approx306.5$ $K$
and standard deviation $\sigma\approx1.6$ $K$, and is slightly asymmetric,
with a longer tail for values below the mean. Overall temperatures
are rather high if compared with normal climatological summer values,
but differences of this order are expected for a perpetual summer
simulation without daily cycle. We analyse the large deviation functions
of the the anomaly of the spatially averaged temperature with respect
to its mean that is, form now on we analyse the observable

\begin{equation}
A(X(t))=T_{\Omega}(X(t))-\mu
\end{equation}
Note that, therefore, the mean of $A(X(t))$ is zero. The large deviation
rate function $A(X(t))$ and of $T_{\Omega}(X(t))$
is the same up to a translation of $\mu$.

From the 1000 year long run we compute the large deviation functions
of the European surface temperature, using the method described in
section \ref{subsec:Estimate-of-large}. First of all we study the
time required to reach the large deviation asymptotic behavior, which
sets the minimum value for the block size $\tau_{b}$. The minimal
requirement is that $\tau_{b}$ is much larger than the autocorrelation
time of the observable. Figure (\ref{fig:a)-Probability-distribution}b)
shows the autocorrelation function of $A(X(t))$

\begin{equation}
R(t)=\frac{\mathbb{E}\left[A(X(t))A(X(0))\right]}{\sigma^{2}}.
\end{equation}
The autocorrelation function is well approximated by a double exponential,
with a first decay on time scale $\tau_{s}\approx$4 days, representative
of the synoptic fluctuations, and a longer decay on a time scale of
about $\tau_{l}\approx$30 days, probably due to the land surface
processes, in particular the dynamics of the soil moisture. The integral
autocorrelation time, computed as $\tau_{c}=\int_{0}^{+\infty}R(t)\,\textrm{d}t$,
results to be $\tau_{c}\approx$ 7.5 days (see Appendix \ref{sec:Direct-estimate-Appendix}
for the details on the computation of $\tau_{c}$). In terms of applications
to the heat wave statistics, an extreme heat wave case can last 1
to 3 months, corresponding to about 4 to 12 autocorrelation times.

\begin{figure}
\begin{centering}
a)\includegraphics[width=0.45\textwidth]{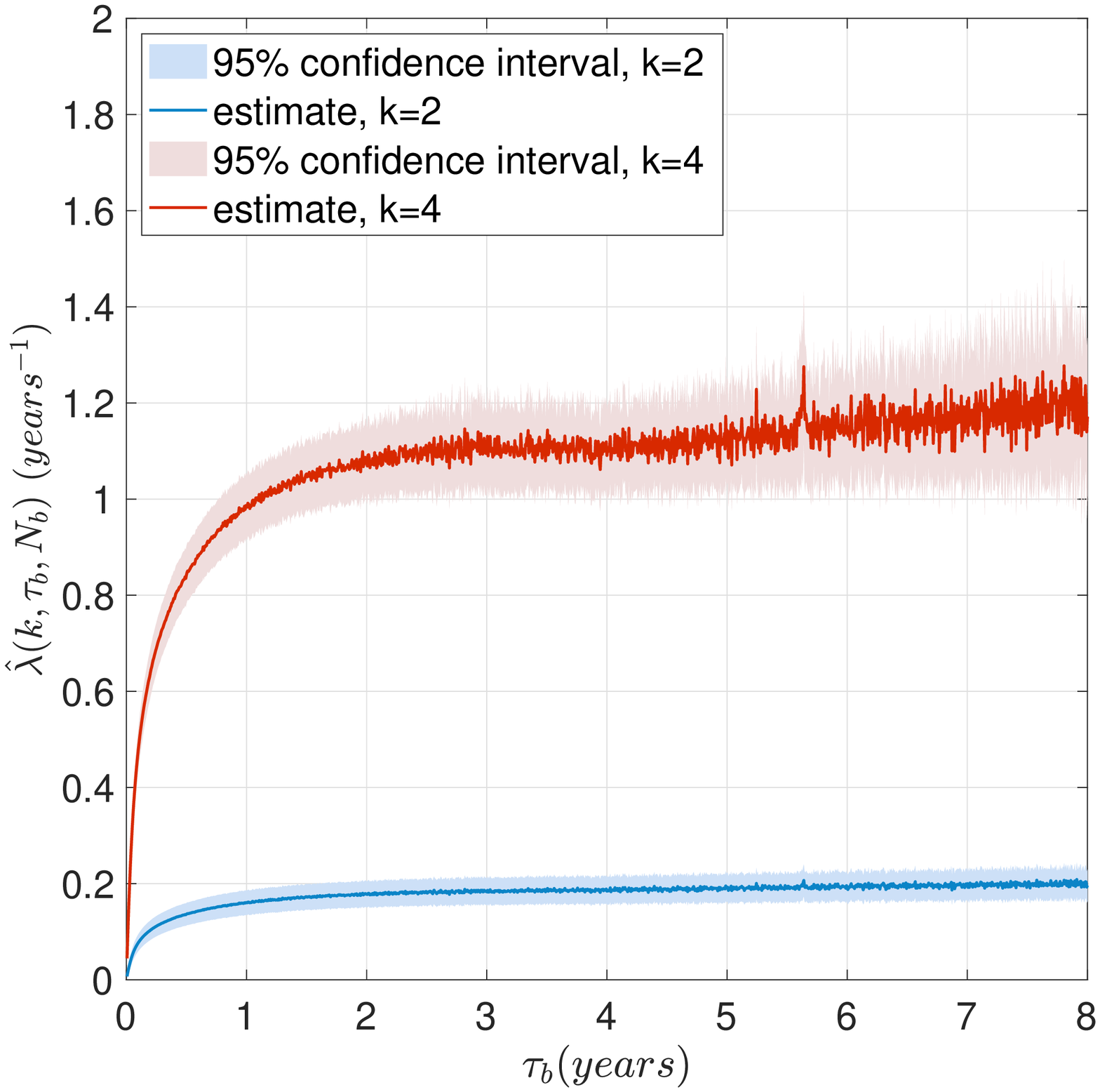} \\ 
b)\includegraphics[width=0.45\textwidth]{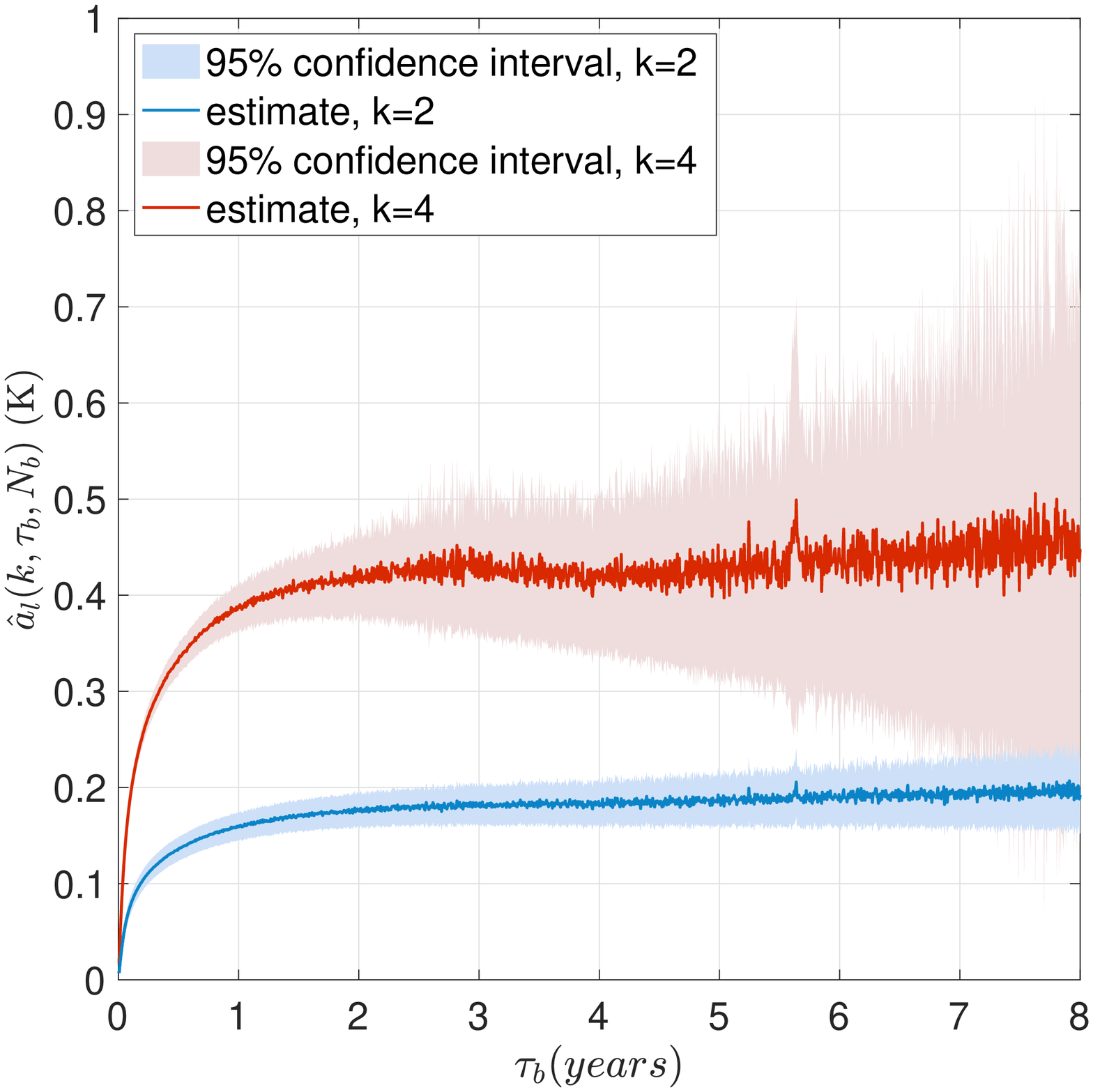} \\
c)\includegraphics[width=0.45\textwidth]{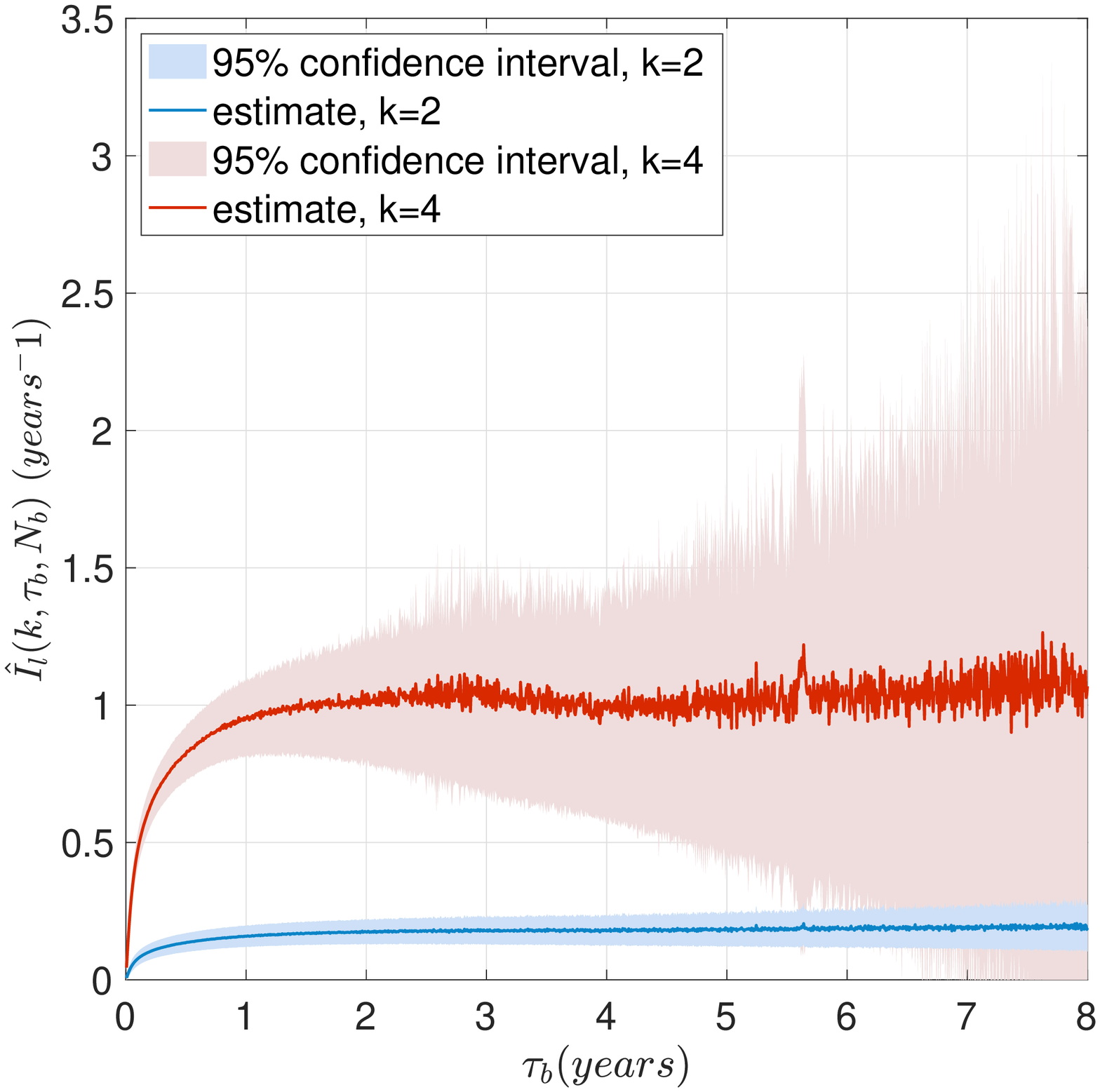}\\
\end{centering}
\caption{For fixed values of $k$, convergence with the averaging time $\tau_{b}$
of the scaled cumulant generating function $\lambda(k)$ (a), the
large deviation estimate of the averaged temperature anomaly $a_{l}=\lambda'(k)$
(b), and the large deviation rate function $I(a)$ (c). For each panel
the two curves show the results for a value of $k$ inside the region
of statistical convergence of both the estimate and its variance ($k=2$,
blue), and a value of $k$ inside the region of statistical convergence
region of the estimate but not of the variance ($k=4$, red).\label{fig:For-fixed-values}}
\end{figure}

The study of the convergence to the large deviation limit shows however
that the time necessary to converge is much longer than that. The
best value of the block size to have proper convergence is $\tau_{b}=3$
$years$ (see Appendix \ref{sec:Direct-estimate-Appendix}). The upper
and lower bounds of the convergence region are estimated at $k_{c}^{+}(\tau_{b},N_{b})=5$
$K^{-1}years^{-1}$ and $k_{c}^{-}(\tau_{b},N_{b})=-2.5$ $K^{-1}years^{-1}$.
Figures \ref{fig:For-fixed-values}a), \ref{fig:For-fixed-values}b)
and \ref{fig:For-fixed-values}c) show the convergence as a function
of $\tau_{b}$ of the scaled cumulant generating function, of its
derivative, and of the rate function, for two values of $k$: one
inside the convergence region of both the estimate and its variance
($k=2$, blue), and one inside the convergence region of the estimate
but not of the variance ($k=4$, red).. They correspond to values
of the temperature anomaly $a$ of respectively $0.21\:K$ and $0.38\:K$.
For example, for $k=2$ the estimated value of $I(a)$ is $0.18\:years^{-1}$,
and in order to reach an estimate of $I(a)$ within a relative error
of 10\% of this value we need $\tau_{b}>1.1\:years$. The analysis
of the convergence thus shows that, at the very least, the averaging
time should be larger than 1 year to consider to be even approximately
in the large deviation asymptotics. Computing a rate function limiting
the averaging time to to 40 days and 90 days gives estimates that
are about 50\% and 65\% of the asymptotic result respectively.

The block averaging approach is based on the idea that, whenever the
processes is sufficiently mixing, one may consider the time average
as an analogous of the sum of $N$ independent variables, $N$ being
of the order of $\tau_{b}/\tau_{c}$. As a rule of thumb the convergence
towards the large deviation rate function would be expected to be
achieved when $N$ is of the order of a few tens. In a case like this,
however, we can see that this reasoning is simplistic. The convergence
here is extremely slow: the minimal $\tau_{b}$ is much larger than
a few times $10\tau_{c}$. The probable reason for this slow convergence
is the slowly decreasing long tail of the correlation function of
the observable. Indeed, the convergence of the large deviation limit
is obtained after about a few years, which is a few times $10\tau_{l}$.

This indicates that the dynamical processes leading to the long decay
time $\tau_{l}=30\,\text{days}$ are connected to the dynamics of
the extreme fluctuations of the time averaged temperature, which correspond
phenomenologically to persistent heat waves. The shorter time scale
$\tau_{s}\approx$4 days is compatible with the classical time scale
of synoptic variability at midlatitudes, essentially the life cycle
of cyclones and anticyclones. The longer time scale $\tau_{l}\approx$30
days is probably due to the low frequency variability of the atmospheric
dynamics and to the atmosphere-surface interactions, in particular
through the soil moisture dynamics. The soil moisture memory is indeed
considered to be a key factor in some of the most extreme heat waves
observed in Europe, like the one of 2003 \cite{Fischer&al2007,Lorenz&al2010,Stefanon&al2012}.

From our analysis it appears that the large deviation limit per se
can not be used to characterize heat waves: due to the presence of
the seasonal cycle in the real world, heat waves are of interest up
to time period of a season, about 90 days, that as we have seen is
far from the time scales for which the large deviation rate function
gives meaningful informations. In a recent paper, \cite{Galfi&al2019}
performed a large deviation analysis of surface temperatures using
the model Puma, which consists in the dynamical core of Plasim, obtaining
faster convergence rates than what we observe. Puma is essentially
Plasim without physical parameterizations, which are substituted by
Newtonian cooling. As a consequence the dynamics in Puma does not
include the range of processes which determine the surface fluxes
of sensible and latent heat, and that lead to memory effects on time
scales longer than the synoptic scale. The faster convergence rates
is very probably due to this aspect. In more realistic setups, from
our analysis it seems unlikely that the statistics of the surface
temperature averaged at regional scale could converge to the large
deviation limit on time scales shorter that a few years, which makes
it not directly relevant for applications.

\begin{figure}
\begin{centering}
(a)\includegraphics[width=0.45\textwidth]{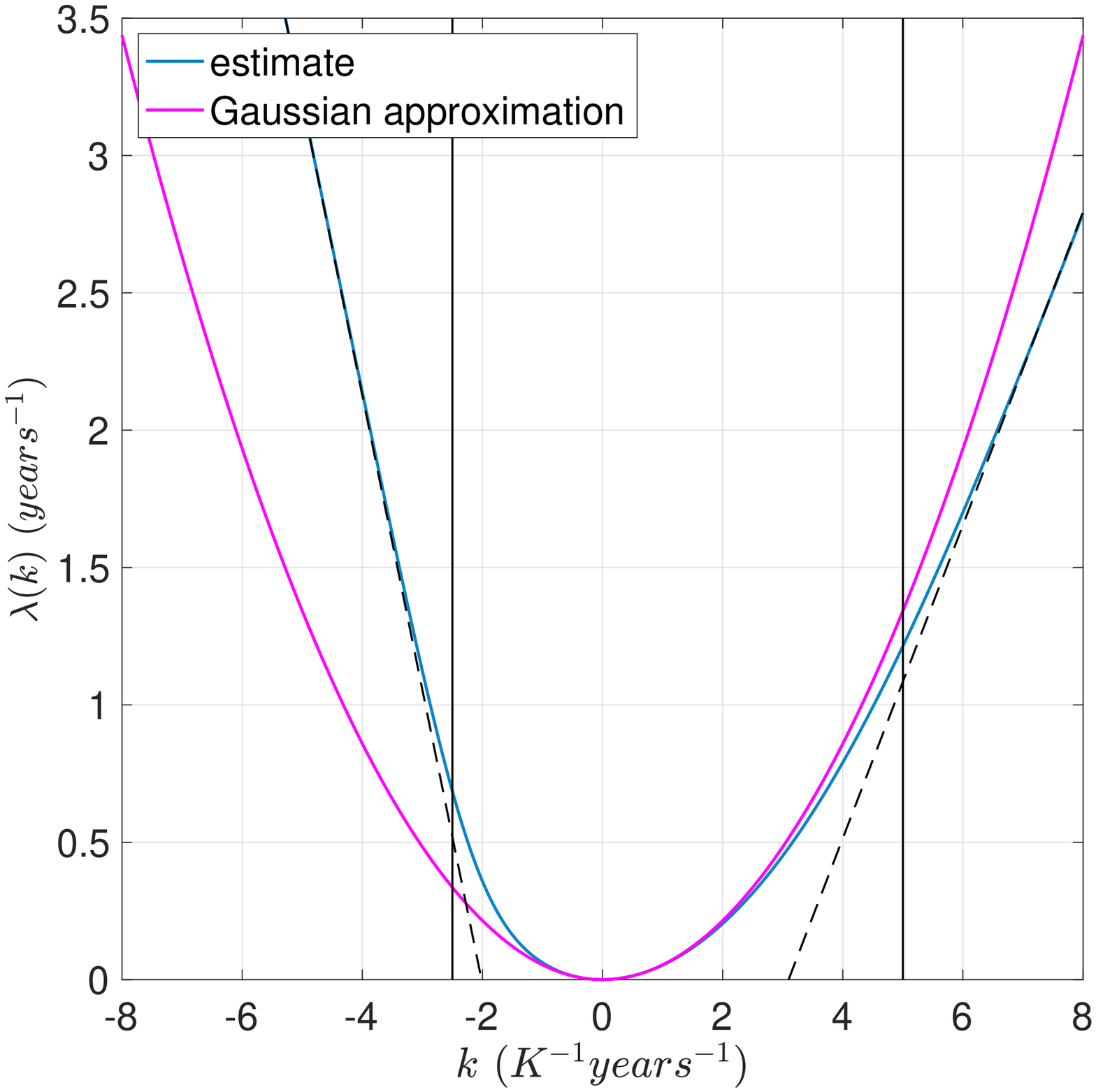} \\
(b)\includegraphics[width=0.45\textwidth]{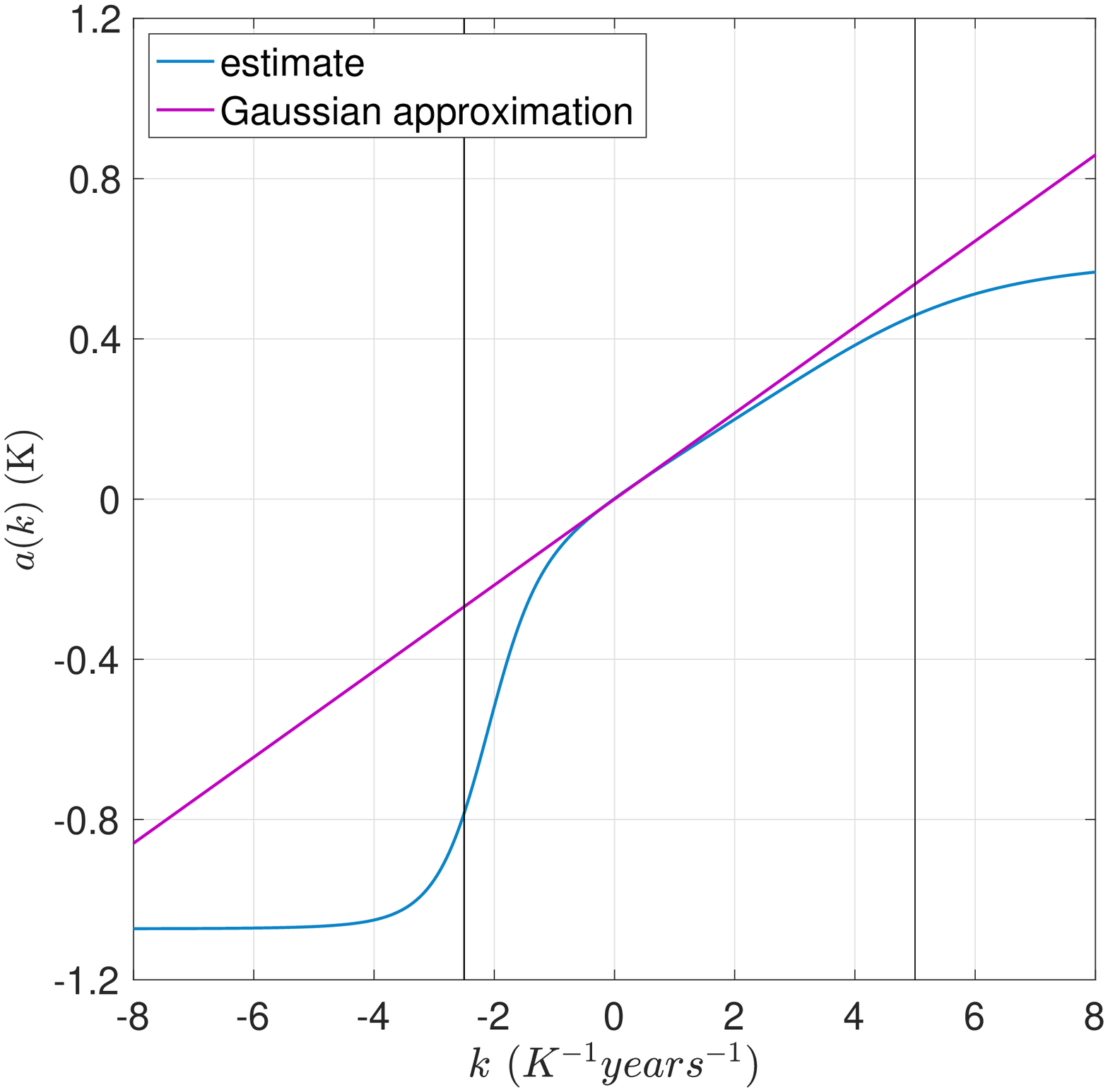} \\
(c)\includegraphics[width=0.45\textwidth]{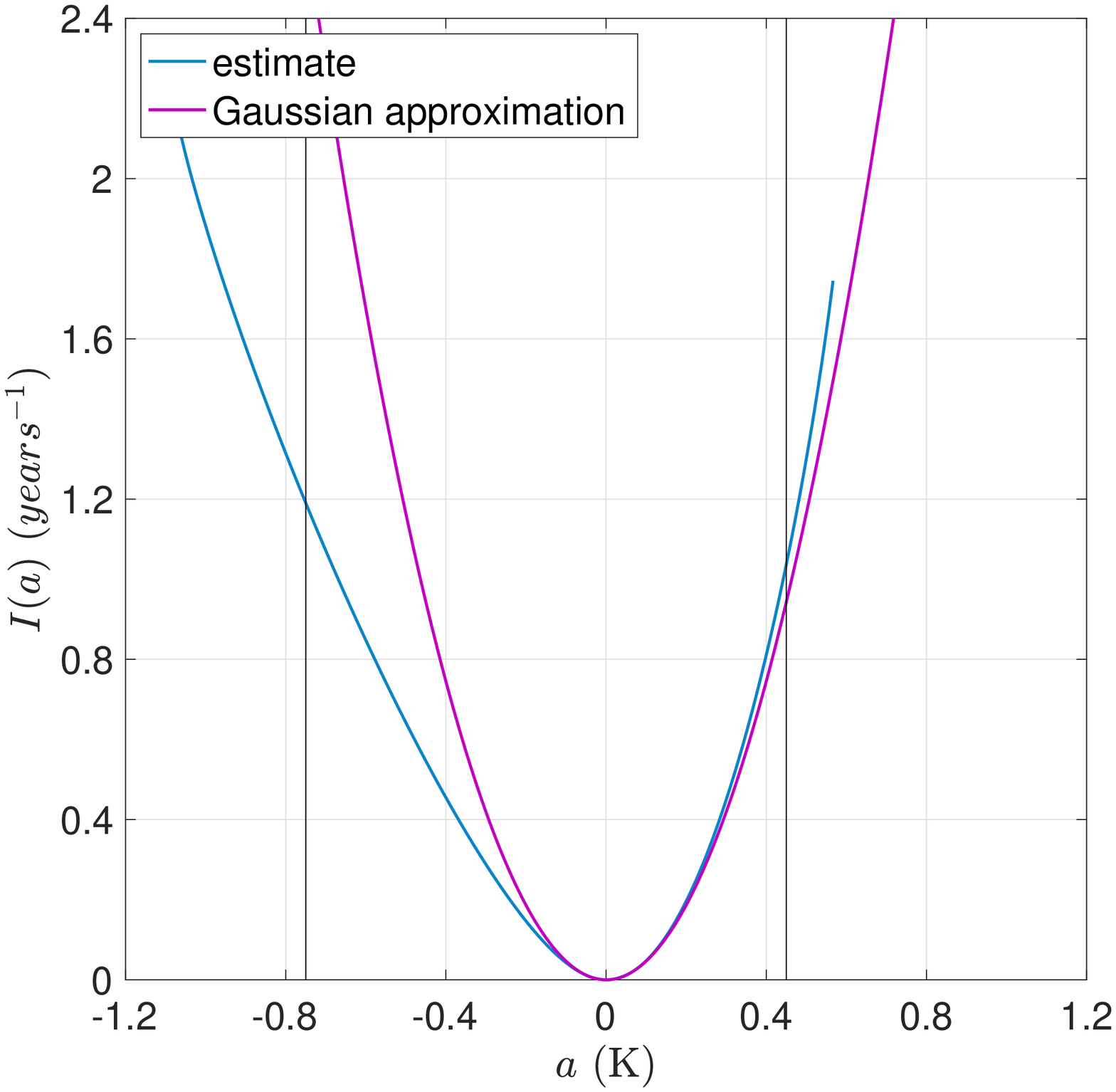}\\
\end{centering}
\caption{For the time averaged European surface temperature, estimates of $\lambda(k)$
the scaled cumulant generating function (panel a)), $a_{l}=\lambda'(k)$
the large deviation estimate of the averaged temperature anomaly (panel
b), and $I(a)$ the large deviation rate (c) versus $k$ or $a$ (blue).
For each cases, the magenta curves represent the Gaussian approximation.\label{fig:Estimate-of-scaled}}
\end{figure}

Figures \ref{fig:Estimate-of-scaled}a), \ref{fig:Estimate-of-scaled}b)
and \ref{fig:Estimate-of-scaled}c) show the best estimates from the
1000 year long run of the scaled cumulant generating function , its
derivative, and the rate function, with $\tau_{b}$=3 years and $N_{b}=$333.
The vertical black lines indicate the boundaries of the convergence
region. The dashed black lines show the artificial asymptotic linear
behavior of the estimate of the scaled cumulant generating function
beyond the convergence region. We see that the large deviation functions
are markedly asymmetric. The rate function is steeper for positive
anomalies than for negative anomalies, and accordingly the scaled
cumulant generating function is larger for positive values than for
negative ones. From the definition (\ref{eq:rate_function}), this
means that large persistent temperature anomalies are much more rare
than cold anomalies with the same magnitude and duration. This can
be quantified at the level of the large deviation rate function. From
Figure \ref{fig:Estimate-of-scaled}c), one can see that the large
deviation rate function value is about the same for a $0.4\,\text{K}$
warm anomaly and for a $-0.7\,\text{K}$ cold anomaly. This means
that the probability of $0.4\,\text{K}$ heat wave that lasts a duration
$T$, decreases exponential with $T$ at a rate of about $1\,\text{y}^{-1}$
($I(0.4)\simeq1\text{y}^{-1}$). The probability of a $-0.7\,\text{K}$
cold spell of duration $T$ decreases at about the same rate ($I(-0.7)\simeq1\text{y}^{-1}$).
However the probability of a $-0.4\,\text{K}$ cold spell of duration
$T$ decreases about $2.5$ times slower ($I(-0.4)\simeq0.4\text{y}^{-1}$). 

Note that \cite{Galfi&al2019} obtained very symmetric large deviation
functions of surface temperature from simulations with the dynamical
core of the same model, and noted that the symmetry was likely unrealistic
and caused by the lack of a proper representation of moist processes
in the model. The fact that in a version of the model that properly
takes into account water phase transitions in the atmosphere the rate
function is rather asymmetric confirms their observation.

The black vertical lines in figures \ref{fig:Estimate-of-scaled}a),
\ref{fig:Estimate-of-scaled}b) and \ref{fig:Estimate-of-scaled}c)
show the convergence regions. The temperature anomalies that can be
properly sampled with $1\,000$ years of data belong to the range
$(-0.8\,\text{°}\text{K},0.4\,\text{°}\text{K})$. As one might expect,
the corresponding values of the probabilities at the two boundaries
of this range are about the same ($I(-0.8)\simeq I(0.4)$). Accordingly
the confidence intervals are larger for negative anomalies which are
less rare. How does this range compare with the Gaussian range? As
can be seen in figures \ref{fig:Estimate-of-scaled}, for positive
anomalies the estimate in the convergence region is still very close
to the Gaussian approximation. For negative anomalies the tail is
markedly less Gaussian. This illustrate that long lasting cold spell
with a return time of about $1\,000$ years can not be studied with
a Gaussian model. In the case of positive anomalies, even 1000 years
of data are not enough to observe the non Gaussian tails of the large
deviation rate function. In order to explore the far tails of the
large deviation functions, one may use rare event algorithms.

\section{Estimate of large deviation functions with rare event algorithms\label{sec:Estimate-of-large}}

\subsection{A large deviation rare event algorithm}

Rare event simulation techniques are numerical tools specifically
dedicated to the computation of rare events in numerical models at
a much smaller computational effort than direct sampling. Such tools
have a long history \cite{kahn_estimation_1951} and have attracted
a growing interest in the last two decades \cite{rubino_rare_2009,bucklew_introduction_2004,giardina_simulating_2011,delmoral_feynmankac_2004}.
The goal of these methods is to make rare events effectively less
rare, thereby improving the efficiency of the statistical estimators.

The method we describe in this paper is a genealogical algorithm originally
proposed by \cite{del_moral_genealogical_2005}, and subsequently
adapted to compute large deviation functions of time averaged observables
in numerical models \cite{giardina_direct_2006,lecomte_numerical_2007}.
In \cite{Ragone&al2018}, we have used this method to study European
heat waves, focusing on seasonal time scales. As we discussed in the
previous section, such times scales are out of the large deviation
asymptotics. Here we show instead how the algorithm can be extremely
useful to compute large deviation functions of climate observables,
overcoming the limitations of direct estimates. In the following we
refer to this algorithm as large deviation algorithm or Giardina-Kurchan-Lecomte-Tailleur
(GKLT) algorithm when used to compute large deviation functions, and
as the Del Moral-Garnier algorithm when used to simulate rare events
outside of the large deviation asymptotics.

We give a general description of the large deviation algorithm; more
details are discussed in \cite{Ragone&al2018} or in the original
papers \cite{giardina_direct_2006,lecomte_numerical_2007}. We perform
simulations of an ensemble of $N$ trajectories $\left\{ X_{n}(t)\right\} $
(with $n=1,2,...,N$) starting from different initial conditions.
The total integration time of the trajectories is denoted $T_{a}$.
We consider an observable $A(X(t))$ of which we want to compute the
large deviations. We define a resampling time $\tau$, and during
the evolution of the system we perform at times $t_{i}=i\tau$ (with
$i=1,2,...,T_{a}/\tau$) a resampling procedure based on the past
values of the observable on the trajectories. At time $t_{i}$ we
assign to each trajectory $n$ a weight $W_{n}^{i}$ defined as
\begin{equation}
W_{n}^{i}=\frac{e^{k^{*}\intop_{t_{i-1}}^{t_{i}}A(X_{n}(t))\,\text{d}t}}{R_{i}}\,\,\,\mbox{with}\,\,\,R_{i}=\frac{1}{N}\sum_{n=1}^{N}e^{k^{*}\int_{t_{i-1}}^{t_{i}}A(X(t))\,\text{d}t},\label{eq:Weight-1}
\end{equation}
where $k^{*}$ is a tuning parameter of the algorithm, whose role
is described in the following. For each trajectory $n$, a random
number of copies of the trajectory are generated at time $t_{i}$.
The expectation value of the number of copies generated by a trajectory
$n$is proportional to its weight $W_{n}^{i}$. Trajectories featuring
large values of the time average of the observable will thus produce
many copies of themselves, while trajectories featuring small values
of the observable will not produce any copies and will effectively
be killed. For practical reasons it is convenient to generate the
copies in such a way to fix the total number of trajectories to be
always exactly $N$ after each resampling, as described in \cite{Ragone&al2018}.
The value of the parameter $k^{*}$ defines how stringent is the selection.
For large values of $k^{*}$ only the trajectories with the very largest
values of the observable will be allowed to generate copies of themselves.
If the system is deterministic, like in the case of climate models,
a small random perturbation is added just after the cloning to each
trajectory, so that copies of the same trajectory will evolve differently.
See \cite{Ragone&al2018} for more details.

Once the final time $T_{a}$ is reached and the simulation is over,
an effective ensemble is reconstructed by removing all the pieces
of trajectories that did not survive until time $T_{a}$. We indicate
with $\mathbb{P}_{0}\left(\left\{ X(t)\right\} _{0\leq t\leq T_{a}}\right)$
the probability of observing a certain trajectory between time 0 and
$T_{a}$ as normally generated by the dynamics of the model, and with
$\mathbb{P}_{k^{*}}\left(\left\{ X(t)\right\} _{0\leq t\leq T_{a}}\right)$
the probability of observing that same trajectory in the effective
ensemble. One can show that

\begin{equation}
\mathbb{P}_{k^{*}}\left(\left\{ X(t)\right\} _{0\leq t\leq T_{a}}\right)\underset{N\rightarrow\infty}{\sim}\frac{e^{k^{*}\int_{0}^{T_{a}}A(X(t))dt}}{\mathbb{E}\left[e^{k^{*}\int_{0}^{T_{a}}A(X(t))dt}\right]}\mathbb{\mathbb{P}}_{0}\left(\left\{ X(t)\right\} _{0\leq t\leq T_{a}}\right),\label{eq:Biased_Path_Approximation-1}
\end{equation}
where $\underset{N\rightarrow\infty}{\sim}$ means that this is true
asymptotically for large $N$ with typical error of order $1/\sqrt{N}$
when evaluating averages over observables. For large positive values
of $k^{*}$, the path measure $\mathbb{P}_{k^{*}}$ is thus tilted
with respect to $\mathbb{P}_{0}$ such that large values of $a=\frac{1}{T}\int_{0}^{T}A(X_{n}(t))\,\mbox{d}t$
will be favored. In order obtain (\ref{eq:Biased_Path_Approximation-1}),
we have used the mean field approximation
\begin{equation}
R_{i}=\frac{1}{N}\sum_{n=1}^{N}e^{k^{*}\int_{t_{i-1}}^{t_{i}}A(X_{n}(t))dt}\underset{N\rightarrow\infty}{\sim}\mathbb{E}\left[e^{k^{*}\int_{t_{i-1}}^{t_{i}}A(X_{n}(t))dt}\right].\label{eq:Mean_Field_Approximation-1}
\end{equation}
The validity of this approximation and the fact that typical relative
errors are of order $1/\sqrt{N}$ have been proved to hold asymptotically
for large $N$ by \cite{delmoral_feynmankac_2004}, for a family
of genealogical algorithms which includes the one adopted here.

The algorithm thus samples very efficiently the tails of the probability
distribution $\rho(a,T)$. Equation (\ref{eq:Biased_Path_Approximation-1})
can be used to compute statistics according to $\mathbb{P}_{0}$ (the
original statistics of the system, what we are interested in) from
an ensemble of trajectories distributed according to $\mathbb{P}_{k^{*}}$
(obtained with a simulation with the algorithm). An estimator of the
expectation value of any quantity $O\left(\left\{ X(t)\right\} _{0\leq t\leq T_{a}}\right)$
based on (\ref{eq:Biased_Path_Approximation-1}) is

\begin{multline}
\mathbb{E}\left[O\left(\left\{ X(t)\right\} _{0\leq t\leq T_{a}}\right)\right]\underset{N\rightarrow\infty}{\sim}\frac{1}{N}\sum_{n=1}^{N}\left(e^{-k^{*}\int_{0}^{T_{a}}A(X_{n}(t))\,\text{d}t}\prod_{i=1}^{T_{a}/\tau}R_{i}\right)O\left(\left\{ X_{n}(t)\right\} _{0\leq t\leq T_{a}}\right),\label{eq:GK_O_estimator}
\end{multline}
where the $X_{n}$ are the $N$ backward reconstructed trajectories
present in the effective ensemble. Note that in (\ref{eq:GK_O_estimator})
there is no assumption of $T_{a}$ being large. Since in the tilted
ensemble large values of $a=\frac{1}{T}\int_{0}^{T}f(x(t))\,\mbox{d}t$
will be more common, the estimation of the tails of $\rho(a,T)$ will
have smaller statistical errors. Tuning $k^{*}$ will allow us to
study different ranges of extreme values in the tails of $\rho(a,T)$.

In the limit of very large $T_{a}$, the algorithm provides a direct
way to compute very efficiently the scaled cumulant generating function.
Using (\ref{eq:Mean_Field_Approximation-1}) 
one obtains that the scaled cumulant generating function at $k=k^{*}$
can be computed as

\begin{equation}
\hat{\lambda}(k^{*})=\lim_{T_{a}\rightarrow\infty}\frac{1}{T_{a}}\sum_{i=1}^{T_{a}/\tau}\log\left[R_{i}\right],\label{eq:GK_lambda_estimator-1}
\end{equation}
with a relative error of order $1/\sqrt{N}$. This is the main output
of the large deviation algorithm in its GKLT formulation \cite{giardina_direct_2006,lecomte_numerical_2007}.
From the scaled cumulant generating function one can then compute
the rate function as described in the previous section.

In typical applications of the large deviation algorithm \cite{giardina_simulating_2011},
the systems under consideration were sufficiently inexpensive to run
that it was possible to perform several experiments with different
values of $k^{*}$ and compute the scaled cumulant generating function
pointwise. With a climate model this is computationally unfeasible.
However, equation (\ref{eq:Biased_Path_Approximation-1}) can be used
to compute a very precise estimate of the scaled cumulant generating
function in a neighborhood of $k^{*}$. Using (\ref{eq:GK_O_estimator})
an estimator of $\lambda(k)$ is

\begin{equation}
\hat{\lambda}(k)=\hat{\lambda}(k^{*})+\lim_{T_{a}\rightarrow\infty}\frac{1}{T_{a}}\log\left[\frac{1}{N}\sum_{n=1}^{N} e^{(k-k^{*})\int_{0}^{T_{a}}A(X_{n}(t))dt}\right].\label{eq:GK_O_estimator-1}
\end{equation}

Equation (\ref{eq:GK_O_estimator-1}) gives a very good estimate of
the scaled cumulant generating function in a neighborhood of $k^{*}$.
This can be seen noting that the typical value of the observable $\frac{1}{T_{a}}\intop_{0}^{T_{a}}A(X(t))dt$
observed in the algorithm statistics in the limit of large $T_{a}$
is, using equation (\ref{eq:Biased_Path_Approximation-1}),

\begin{multline}
\mathbb{E}_{k^{*}}\left[\lim_{T_{a}\rightarrow+\infty}\frac{1}{T_{a}}\intop_{0}^{T_{a}}A(X(t))\,\mbox{d}t\right]=\lim_{T_{a}\rightarrow+\infty}\mathbb{E}\left[\left(\frac{1}{T_{a}}\intop_{0}^{T_{a}}A(X(t))\,\text{d}t\right)\frac{e^{k^{*}\int_{0}^{T_{a}}A(X(t))\,\text{d}t}}{\mathbb{E}\left[e^{k^{*}\int_{0}^{T_{a}}A(X(t))\,\text{d}t}\right]}\right]=\lambda^{'}(k^{*}).
\end{multline}
Since in the large deviation regime the average and the most probable
value coincide as a first approximation, this means that the time
average of the observable in the effective ensemble will fluctuate
around a typical value given by the derivative of the scaled cumulant
generating function evaluated in $k^{*}$. This is exactly the range
of fluctuations that are needed in order to compute correctly the
large deviation function in a neighborhood of $k^{*}$ \cite{Rohwer&al2015}.
Effectively tilting the trajectory probability density shifts the
center of the convergence region of the estimator around $k=k^{*}$.
It is therefore possible to perform a few experiments with different
values of $k^{*}$, compute the estimate of $\lambda(k)$ for the
neighborhoods of the values of $k^{*},$ and then join the estimates
to reconstruct $\lambda(k)$ piecewise. This allows to explore the
tails of the scaled cumulant generating function (hence of the rate
function) with a huge gain in terms of computational cost with respect
to direct estimation methods.

\subsection{Analysis of large deviations of Europe surface temperature in a climate
model with the large deviation algorithm}

We demonstrate here the performances of the large deviation algorithm
applied to the climate model Plasim to compute the large deviation
functions of the European surface temperature. We use the direct estimate
obtained with the 1000 years long control run as a benchmark and show
that with the algorithm it is possible to compute the same values
of the large deviation functions with a smaller computational cost.

The experiments are carried out for different values of $k^{*}$ with
$N=128$ trajectories, each run for a total time $T_{a}=$800 days.
Each experiment has thus a total computational cost of about 284 years.
The initial conditions of the trajectories are taken from the control
run, spaced by a few years from each other to ensure statistical independence.
The first 80 days of simulations are considered as a transient to
reach statistical equilibrium, and the statistical analysis is performed
on the last 720 days. The resampling time is set at 8 days, as in
\cite{Ragone&al2018}. The choice of the resampling time is determined
by how trajectories starting from the same initial condition separate
in time after the addition of a small random perturbation. Heuristically
it is expected that a good choice should be $\tau$ to be of the order
of $\tau_{c}$. If the resampling time is much smaller than the autocorrelation
time of the process, the trajectories do not have time to separate
enough and useless resampling will increase the variance. If on the
contrary the resampling time is much larger than the autocorrelation
time, the trajectories will fall back to the typical states, and the
importance sampling efficiency will be lost. Tests with simpler systems\textbf{
}\cite{Lestang&al2018}\textbf{ }have shown that the precise value
of $\tau$ does not affect the results, as long as it is of the order
of the autocorrelation time. The trajectories are perturbed after
cloning by adding a small random field to the surface pressure, as
described in \cite{Ragone&al2018}.

\begin{figure}[t]
\begin{centering}
(a)\includegraphics[height=0.45\textwidth]{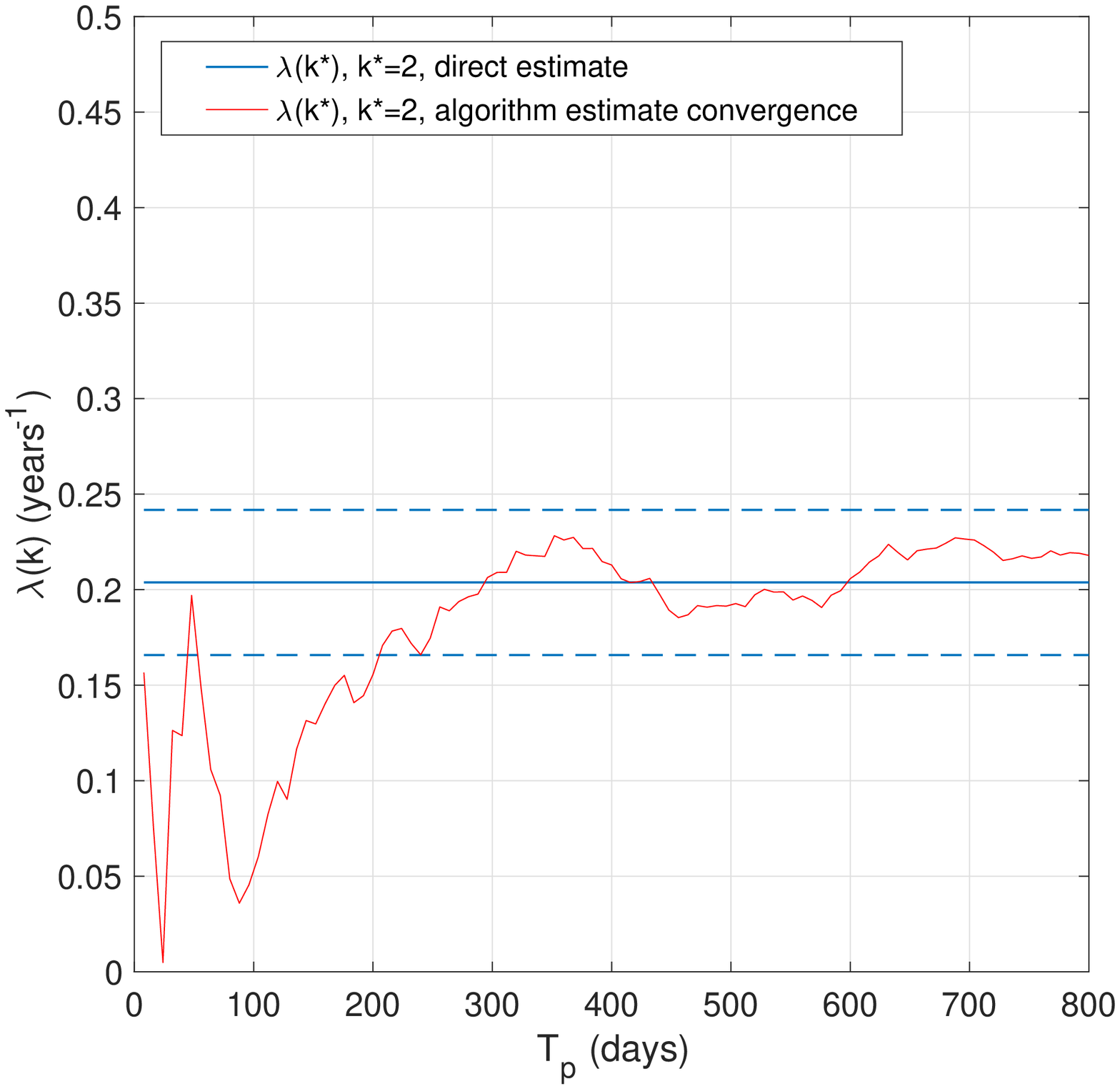}\\
(b)\includegraphics[height=0.45\textwidth]{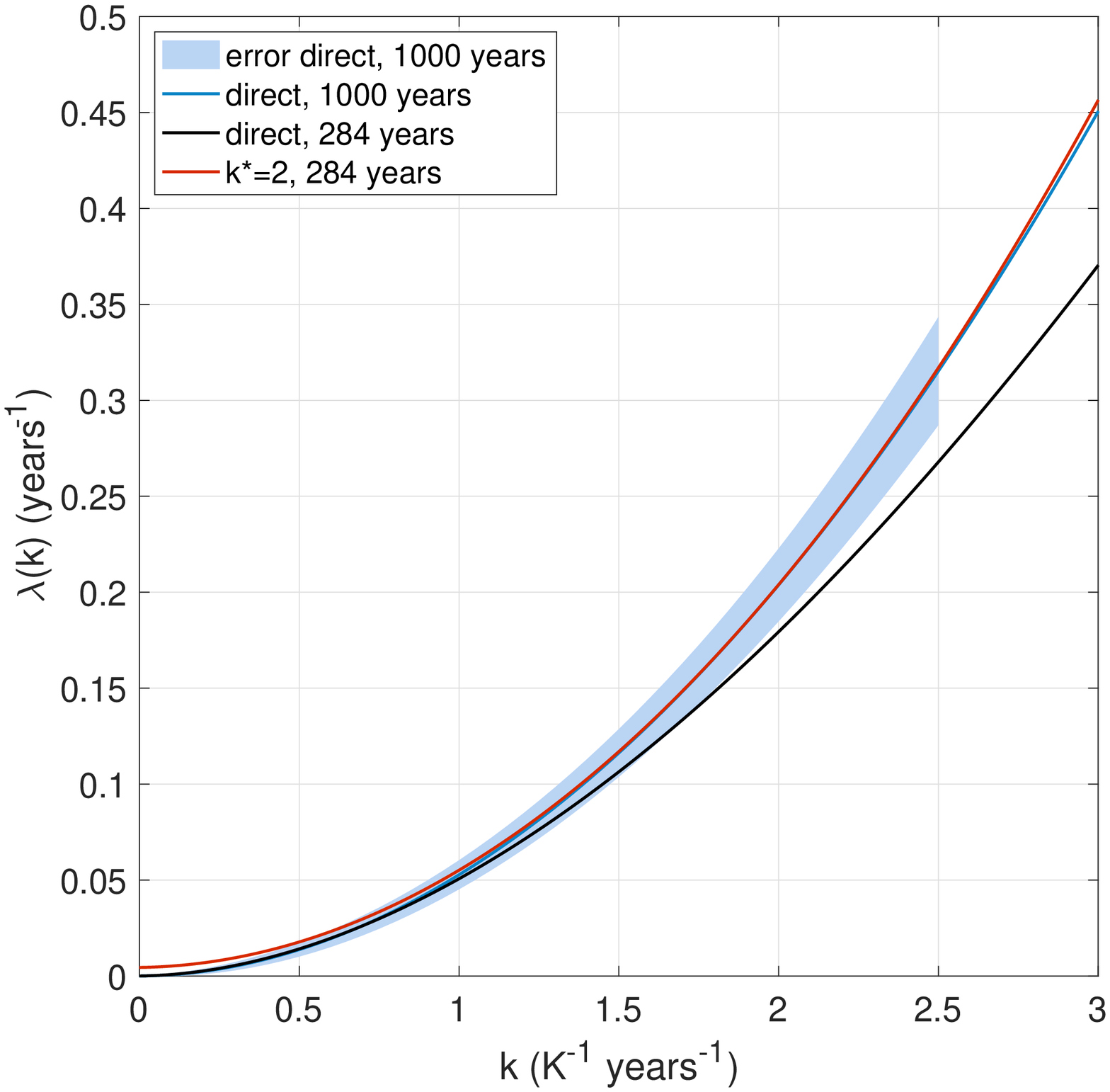}\\
\end{centering}
\caption{(a) Estimate of $\lambda(k^{*})$ with $k^{*}=2$. The blue solid
line indicates the reference value obtained from the control run,
with the blue dashed lines indicating the 95\% confidence interval.
The red line shows the convergence of the estimate obtained with the
algorithm as a function of the length of the simulation $T_{p}$.
From simulations of 300 days or longer the estimate of $\lambda(k^{*})$
oscillates stably well inside the 95\% confidence interval of the
direct estimate. (b) Direct estimate of the scaled cumulant generating
function from the 1000 years long control run (blue), direct estimate
from the control run using only 284 years (black) and estimate with
the algorithm with $k^{*}=2$ (red). \label{fig:(a)-k2}}
\end{figure}

We first consider $k^{*}=2$, a value for which the direct estimate
of the large deviation rate function with the 1000 years control run
converges. Figure \ref{fig:(a)-k2} shows the convergence of the estimate
obtained with the algorithm as a function of the length of the simulation,
that is the quantity

\begin{equation}
\hat{\lambda}(k^{*})=\frac{1}{T_{p}}\sum_{i=1}^{T_{p}/\tau}\log\left[R_{i}\right],
\end{equation}

with the partial simulation length going from $T_{p}=0$ to $T_{p}=T_{a}=800$
days. From simulations of 300 days or longer the estimate of $\lambda(k^{*})$
oscillates stably well inside the 95\% confidence interval of the
direct estimate. Note that $T_{p}=300$ days corresponds to a total
computational cost of about 107 years. The convergence speed depends
on the value of $k^{*}$, hence the choice of considering in general
a length of $T_{p}=T_{a}=800$ days, to stay on the safe side. The
estimate of $\lambda(k^{*})$ using the algorithm is a sum of $T_{a}/\tau$
values, each for each resampling. We can associate to the estimate
an error as the standard deviation associated to such sum, divided
by the square root of $T_{a}/\tau$. In this case we have that for
$k^{*}=2$ the rare event algorithm estimate gives $\lambda(k^{*})=0.218\pm0.025$
against a direct estimate of $\lambda(k^{*})=0.204\pm0.038$.

The data from the large deviation algorithm can also be used also
to estimate $\lambda(k)$ in a neighborhood of $k^{*}$. Figure \ref{eq:GK_lambda_estimator-1}b
shows the estimate of $\lambda(k)$ for $0<k<3$ obtained with the
experiment with $k^{*}=2$ , using equation (\ref{eq:GK_O_estimator-1}),
compared with the direct estimate obtained from a 1000 years of control
run and with a direct estimate obtained using only 284 years of the
control run. We can see that the estimate of the large deviation algorithm
perfectly coincides with the direct estimate from the long run in
its convergence region. By contrast the direct estimate for short
control run performs poorly for $k>1$, due to the lack of statistics.
In particular, for $k>1.75$ the estimate is outside of the confidence
interval of the direct estimate from the long control run, so that
the short control run estimate is clearly wrong for those values of
$k$. As already said, the large deviation algorithm essentially shifts
the center of the convergence region from $k=0$ to $k=k^{*}$ . Consequently,
the best agreement with the benchmark is obtained for $k=2$, while
the agreement is slightly worse close to $k=0$. We can conclude that
indeed the large deviation algorithm outperforms the estimation of
the scale cumulant generating functions in range of $k$ centered
around the value used in the algorithm.

One can use the algorithm to extend the estimates to values of $k$
that are outside the convergence region of the long control run, for
which we would need an extremely long simulation in order to use the
direct sampling method. The idea is to perform a series of $N_{k}$
experiments with different values of $k^{*}=k_{i}^{*}$, with $i=1,2,...,N_{k}$,
each providing a local estimate $\hat{\lambda}_{i}(k,\tau_{b},N_{b})$
around $k_{i}^{*}$. We then reconstruct the scaled cumulant generating
function, for instance by piecewise linear approximations $\hat{\lambda}(k)=\alpha_{i}\hat{\lambda}_{i}(k)+(1-\alpha_{i})\hat{\lambda}_{i+1}(k),\,\,\,\mbox{with}\,\,\,k_{i}^{*}\leq k<k_{i+1}^{*}$
where $\alpha_{i}=(k_{i+1}^{*}-k)/(k_{i+1}^{*}-k_{i}^{*}).$ In order
to make a proper analysis we have to add an uncertainty to the estimates.
However, in the case of the large deviation algorithm we have no rigorous
results on the range of convergence of the estimator and of its variance.
Empirical analysis shows that the estimates of the scale cumulant
generating function obtained with the large deviation algorithm suffers
from the same linearization problem as the direct estimate, only with
the center of the convergence region shifted around $k=k^{*}$. Therefore,
we can estimate the error of the scale cumulant generating function
and its range of convergence by simply mimicking what we did in the
case of the direct method.

\begin{figure}
\begin{centering}
(a)\includegraphics[width=0.45\textwidth]{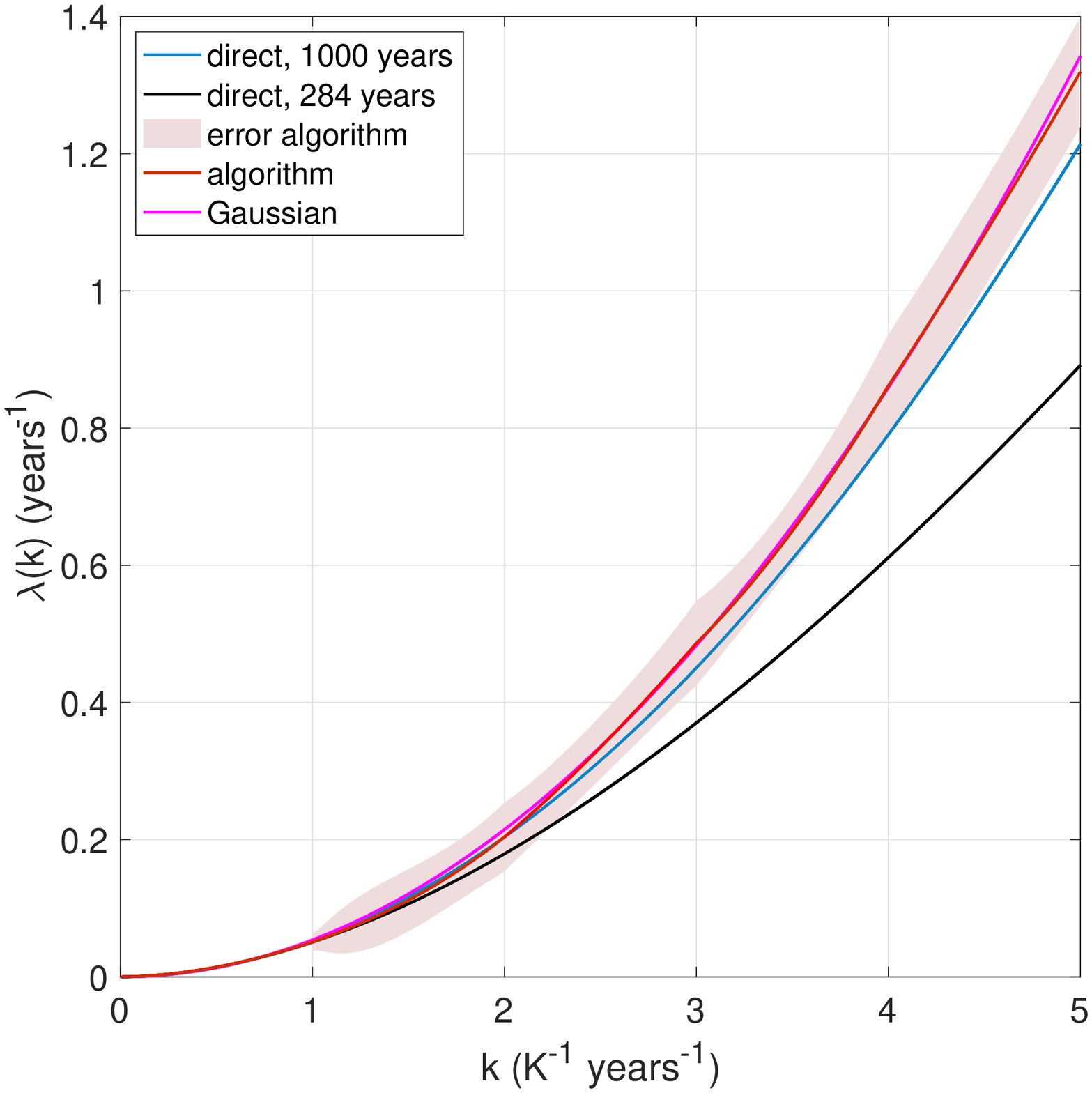}\\ 
(b)\includegraphics[width=0.45\textwidth]{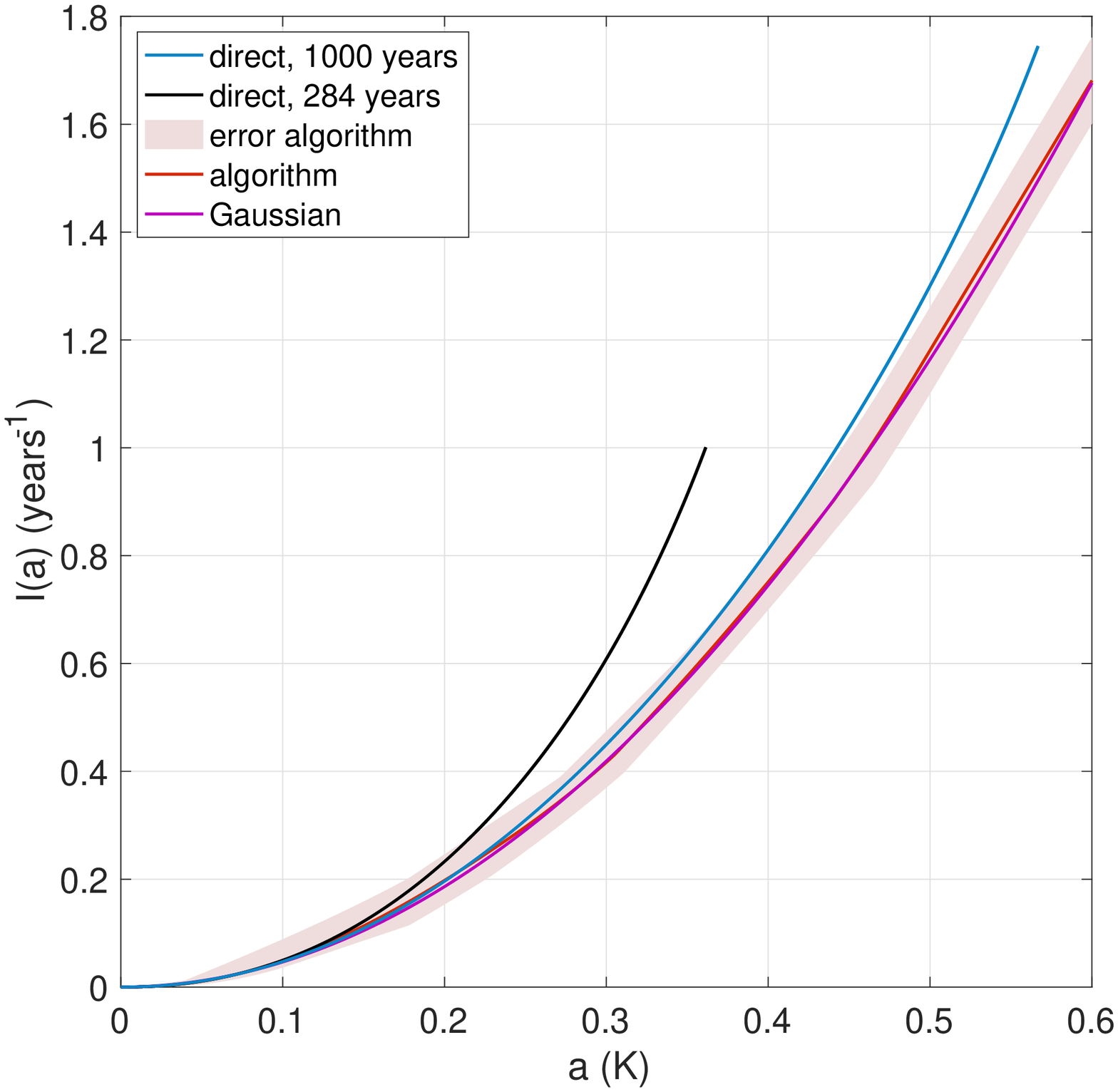}\\
\end{centering}
\caption{(a) Direct estimate of the scaled cumulant generating function from
the 1000 years control run (blue), direct estimate from the 284 years
control run (black) and estimate obtained combining the results obtained
with the algorithm with $k^{*}=2,3,4$ (red). (b) The same for the
rate function. \label{fig:(a)-reconstruction}}
\end{figure}

Figure \ref{fig:(a)-reconstruction} shows the scaled cumulant generating
function reconstructed in this way using three experiments with $k^{*}=2,3,4$.
For $0\leq k<1$ we have used only the 284 years direct estimate,
for $1\leq k<2$ we have used the 284 years direct estimate and the
$k^{*}=2$ estimate, and for $k\geq2$ we have used the method described
above. We note that the 284 year direct estimate is clearly wrong
for large values of $k$. Moreover, for $k>3.5$, the 1000 years control
run estimate is outside the error bar of the large deviation algorithm
estimate, which is the most reliable in this range. This is consistent
with last section results that concluded that the 1000 years control
run estimate is unreliable for $k>2.5$. We would have needed much
more than 1000 years of data in order to have a reliable direct estimate
for those values of $k$.\\

Somehow surprisingly, looking at figure \ref{fig:(a)-reconstruction},
we learn that the large deviation function for warm anomalies is extremely
Gaussian. Relying only on direct estimates without performing a proper
convergence analysis, we may have led us to the wrong conclusion that
1000 years of data would have been sufficient to go beyond the Gaussian
regime. Instead, the large deviation functions are Gaussian well beyond
the convergence region of the direct estimate even for such a long
run. If one is interested in computing the non Gaussian part of the
tail the use of rare event sampling techniques seem thus of vital
importance. A piecewise reconstruction of the large deviation functions
with the algorithm can seem computationally expensive, as one has
to perform several experiments for different values of $k^{*}$. However,
in this case the computational cost grows linearly with the size of
the range of $k$ one wants to explore, while in the case of the direct
estimate it grows exponentially, making a proper analysis impossible.

\section{Using the large deviation algorithm for extreme heat waves \label{sec:Extreme-heat-waves}}

As we have discussed, the time scales required to reach the large
deviation asymptotics are too long for the large deviation rate functions
to be relevant to discuss seasonal heat waves or cold spells. However,
the rare event algorithm itself \cite{del_moral_genealogical_2005}
does not need $T_{a}$ to be in the large deviation asymptotics. In
\cite{Ragone&al2018} we have exploited this property to show how
the algorithm could be used to sample extremely rare heat waves with
return times up to millions of years, with computational costs two
to three orders of magnitude smaller. Here we recall the main results
of \cite{Ragone&al2018}, connecting them more explicitly with the
previous sections.

We are interested in computing the tail of the distribution of the
European surface temperature time averaged over a time $T$ of interest
for applications. The strongest heat waves are characterized by persistence
at scales from sub-seasonal to seasonal (between a few weeks and 3
months). This value of $T$ is more than one order of magnitude smaller
than the values we have considered in the previous section. Consequently,
the value of the algorithm parameter $k^{*}$ must be chosen carefully.
The value of the parameter $k^{*}$ determines the most probable value
of the fluctuations observed in a simulation with the algorithm, independently
of the value of $T$. When we lower $T$ to study time scales more
realistic for heat waves, the width of $\rho(a,T)$ increases substantially.
This means that the values of the fluctuations $a$that one obtains
with the values of $k^{*}$ that used in the previous section to study
the large deviation limit will not correspond to very rare events
for smaller values of $T$.\textbf{ }Therefore we need to use values
of the biasing parameter $k^{*}$ much larger than those used in the
large deviation limit, to reach fluctuations that are rare for the
new values of $T$.

The choice of the range of values of $k^{*}$ to be used depends on
the value of the averaging time $T$ and on the range on temperature
anomalies $a$ that one wants to analyze. In the large deviation limit,
the relation $\mathbb{E}_{k^{*}}[a]=\lambda'(k^{*})$ can be used
to choose the value of $k^{*}.$ For smaller $T$ the relation does
not hold anymore. However, one can still think to use it to get a
rough estimate at least of the order of magnitude of the value of
$k^{*}$ necessary to target a certain range of fluctuations. Since
the value of $k^{*}$ to be used is very large, a direct estimate
of $\lambda(k^{*})$ and thus $\lambda'(k^{*})$ is not be available,
for the reasons discussed in the previous sections. One could use
then the Gaussian approximation of the scaled cumulant generating
function to have at least the order of magnitude of $k^{*}$ necessary
to observe fluctuations of order $a$, obtaining $k^{*}\approx a/\left(2\tau_{c}\sigma^{2}\right)$.
This is a very crude approximation, but it is still better than a
blind guess.

\begin{figure}
\begin{centering}
(a)\includegraphics[width=0.45\textwidth]{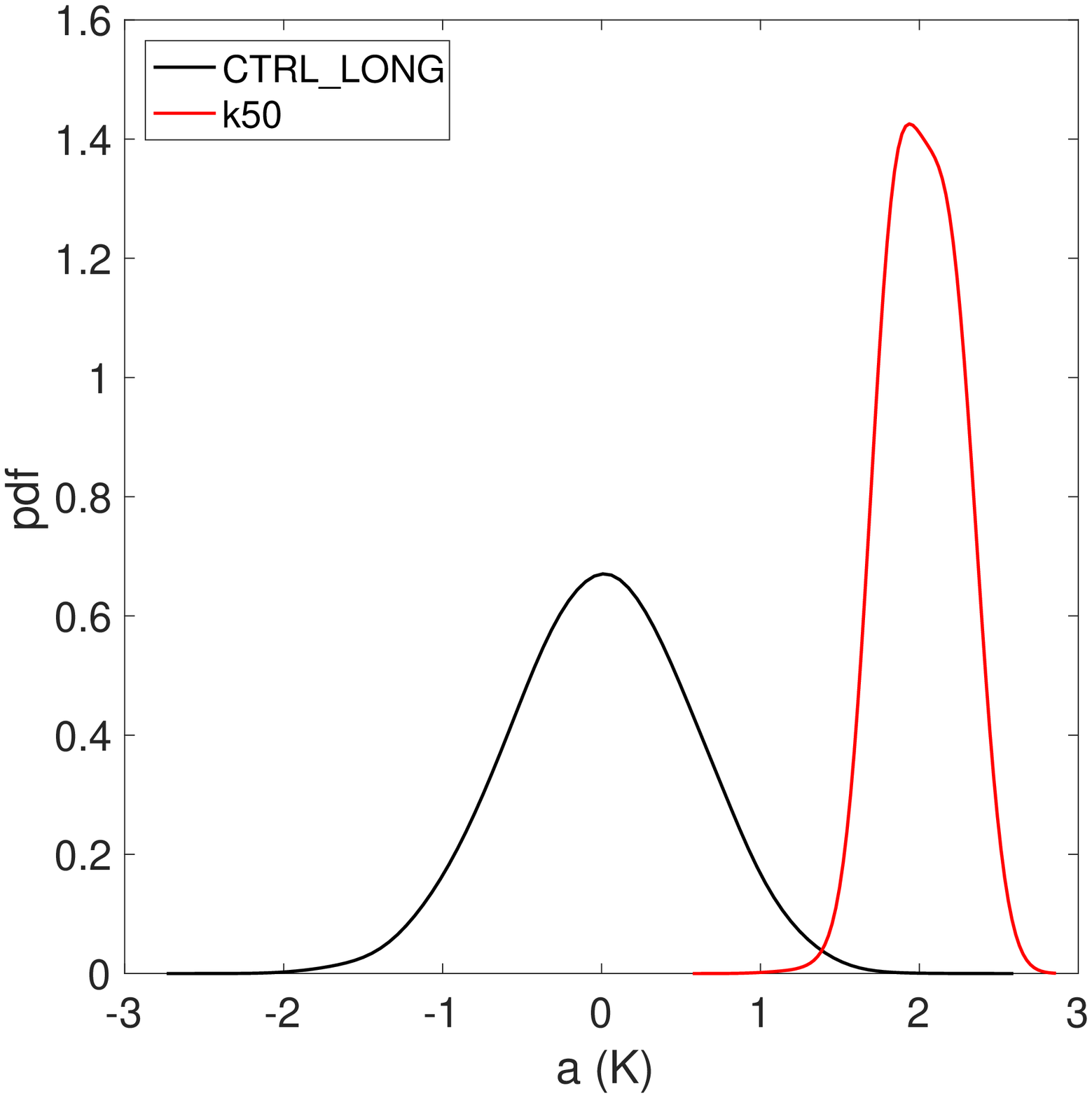}\\ 
(b)\includegraphics[width=0.45\textwidth]{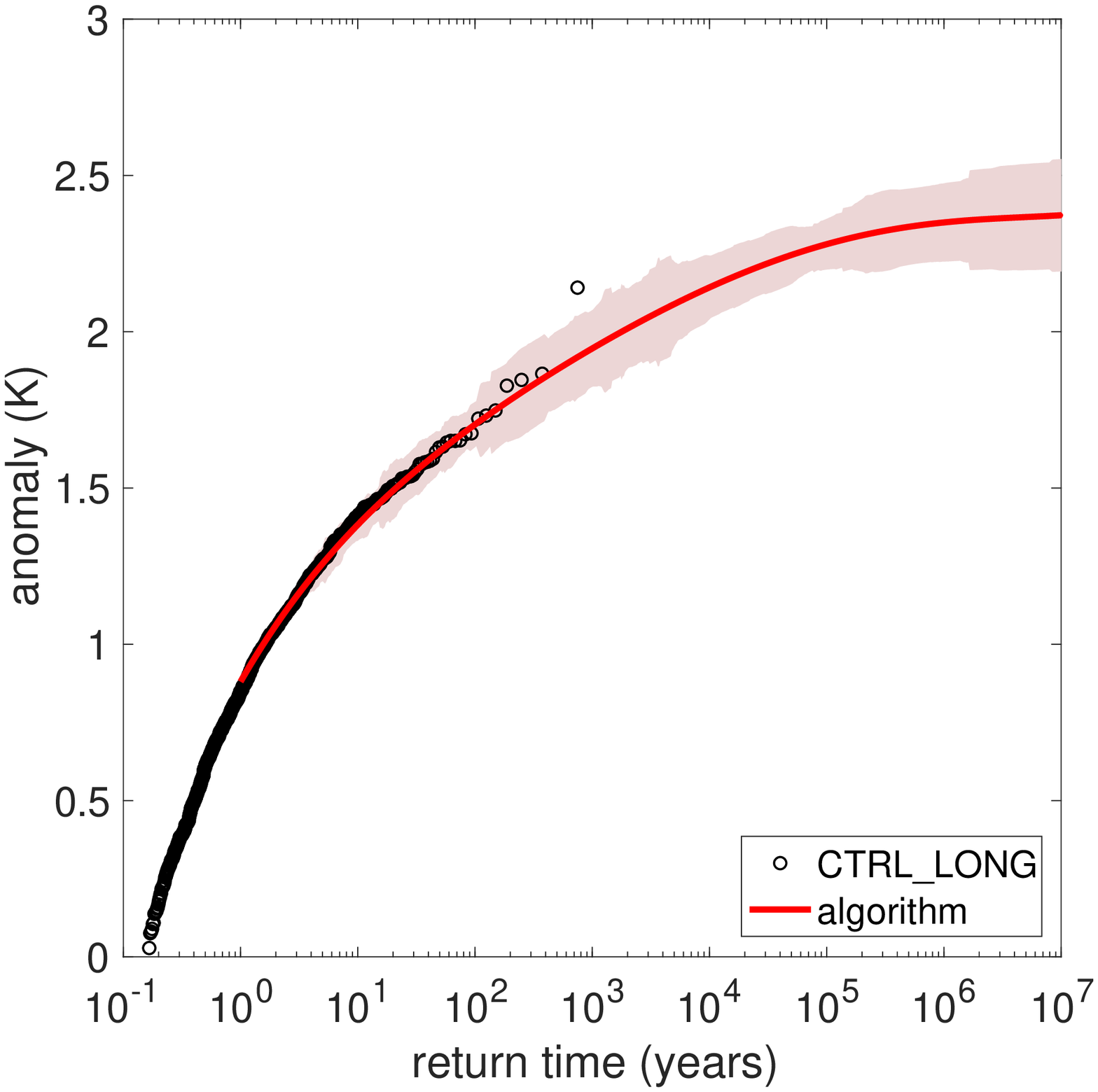}\\
\end{centering}
\caption{(a) Probability distribution function of the 90 days average European
surface temperature, from the control run (black) and the $k^{*}=50$
experiment (red). (b) Return time of the 90 days average European
surface temperature, from the control run (black) and the experiments
with $k^{*}=10,20,40,50$ experiments (red). \label{fig:Pdf-(a)-and} }
\end{figure}

As a test case, we study $T=90$ days. The domain of $k$ over which
the scaled cumulant generating function is known, from the long control
run, with an acceptable degree of accuracy is too narrow to include
values of the derivative of the scaled cumulant generating function
above $\sim1\:K$. The black curve of figure \ref{fig:Pdf-(a)-and}a
shows the probability distribution of the 90 days averaged European
surface temperature estimated from the 1000 years long control run.
We can see that a threshold on 1 K does not actually select extremely
rare events, confirming the discussion above. In the following we
study heat waves for which the time averaged Europe surface temperature
during $T=90$ is larger than $a=2\,\text{K}$, that are extremely
rare. Using the Gaussian approximation of the scaled cumulant generating
function, we infer that the value of $k^{*}$ for which fluctuations
of order $a=2\:K$ are typical is $k\approx19$. Since this is an
order of magnitude argument, we consider for our experiments four
different values, $k^{*}=10,20,40,50$.

The resampling time $\tau$ is kept at 8 days, while the length of
the simulations and the number of trajectories are changed with respect
to the case of the large deviation limit. While in the case of the
computation of the large deviation function it was crucial to wait
for the system to reach statistical equilibrium, and thus $T_{a}$
had to be very large, in this case it is sufficient to take $T_{a}$
at least larger than 90 days to resolve both the transient leading
to the extreme heat wave and the heat wave itself. Since the value
of $k^{*}$ is be much larger than in the large deviation case, it
is necessary to have a larger number of trajectories, in order to
avoid ending up with an effective ensemble populated by the clones
of just one trajectory. We set $T_{a}=128$ days and $N=512$ trajectories.
Each experiment has thus a computational cost of about 182 years.

Figure \ref{fig:Pdf-(a)-and}a) shows the probability distribution
function of the 90 day average of the European surface temperature
for the experiment with $k^{*}=50$ compared with the probability
distribution function of the control run. We can see that indeed the
distribution of the anomalies of the 90 days averaged temperature
is heavily shifted towards positive values. For a $2\,\text{K}$ threshold,
in the $k^{*}=50$ experiment the system is in heat wave conditions
in about 50\% of the trajectories. Thanks to importance sampling,
we can hugely improve the estimate of the statistics of the events
belonging to the tail of the original distribution.

The return time of extreme events is an important characterization
of extremes. Figure \ref{fig:Pdf-(a)-and}b) shows the return time
of the anomalies of the 90 days temperature, estimated from the 1000
years long control run (black). With direct sampling clearly we can
not estimate correctly events with return time larger than a few centuries.
Thanks to the rare event algorithm, we can extend the estimate of
the return time curve to much rarer events. The return time can be
computed from the output of the rare event algorithm as described
in details in \cite{Ragone&al2018} and \cite{Lestang&al2018}.
The red line in figure \ref{fig:Pdf-(a)-and}b) has been obtained
by computing return time functions from the experiments with the algorithm
(the case $k^{*}=20$ and $k^{*}=40$ repeated twice with different
sets of initial conditions to improve the statistics) and averaging
the results in the areas of overlap. Using several experiments with
different values of $k^{*}$ and different sets of initial conditions
helps to improve the statistics and to obtain a better estimate. The
total computational cost of the experiments is of about 1090 years,
basically the same of the control run. We can see that the return
time curve obtained with the large deviation algorithm overlaps with
the upper part of the curve given by the control run (confirming the
correctness of the procedure), but extends to much larger values of
the return time. With the algorithm we can compute return times up
to $10^{6}-10^{7}$ years with a total computational cost of the order
of $10^{3}$ years. There is thus a gain of more than three orders
of magnitude in the sampling efficiency. 

This result has two implications. First, in this way it is possible
to observe ultra rare events that could never be observed in a direct
numerical simulation, unless one employs an unrealistic amount of
computational resources. Second, and possibly more importantly, the
quality of the statistics of rare events in general is greatly improved.
For example, in 1000 years in the control run there is only one event
with temperature in excess of 2 K during the 90 days period. With
the rare event algorithm instead, we have access to several hundreds
of them even considering only one of the experiments, at a fraction
of the computational cost. This improvement of the statistics allows
to perform studies of extreme events that are out of reach with standard
direct sampling.

We can for example compute composite maps of the surface temperature
anomalies and of the 500 hPa geopotential height anomalies, conditional
on the occurrence of an heat wave. This kind of composite statistics
is sometimes used in the study of climatic extremes, in order to detect
the typical dynamical patterns connected to the extremes of interest.
For example, \cite{Stefanon&al2012} provided a classification of
European heat waves by performing a cluster analysis of composites
of heat wave events, although the condition of occurrence of an heat
wave in their case was defined in a more complex way than in this
study. How can this type of analysis be performed with data obtained
with the large deviation algorithm? Let us write either the surface
temperature or the 500 hPa geopotential height anomaly field as $O\left(X(t)\right)$,
and the condition of occurrence of the heat wave as $\frac{1}{T}\int_{0}^{T}A(X(t))dt>a$,
with $T$=90 days and $a$= 2 K. The conditional expectation value
can be computed as

\begin{equation}
\mathbb{E}\left[O\left(X(t)\right)\left|\,\frac{1}{T}\intop_{0}^{T}A\left(X(t)\right)dt>a\right.\right]=\frac{\mathbb{E}\left[O\left(X(t)\right)\Theta\left(\frac{1}{T}\int_{0}^{T}A\left(X(t)\right)dt-a\right)\right]}{\mathbb{E}\left[\Theta\left(\frac{1}{T}\int_{0}^{T}A\left(X(t)\right)dt-a\right)\right]},\label{eq:conditional}
\end{equation}
where $\Theta(\cdot)$ is the Heaviside function. Empirical estimators
of conditional expectation values can be easily extended to the case
of the tilted trajectory probability by using (\ref{eq:GK_O_estimator})
at numerator and denominator of (\ref{eq:conditional}).

\begin{figure}
\begin{centering}
(a)\includegraphics[width=0.45\textwidth]{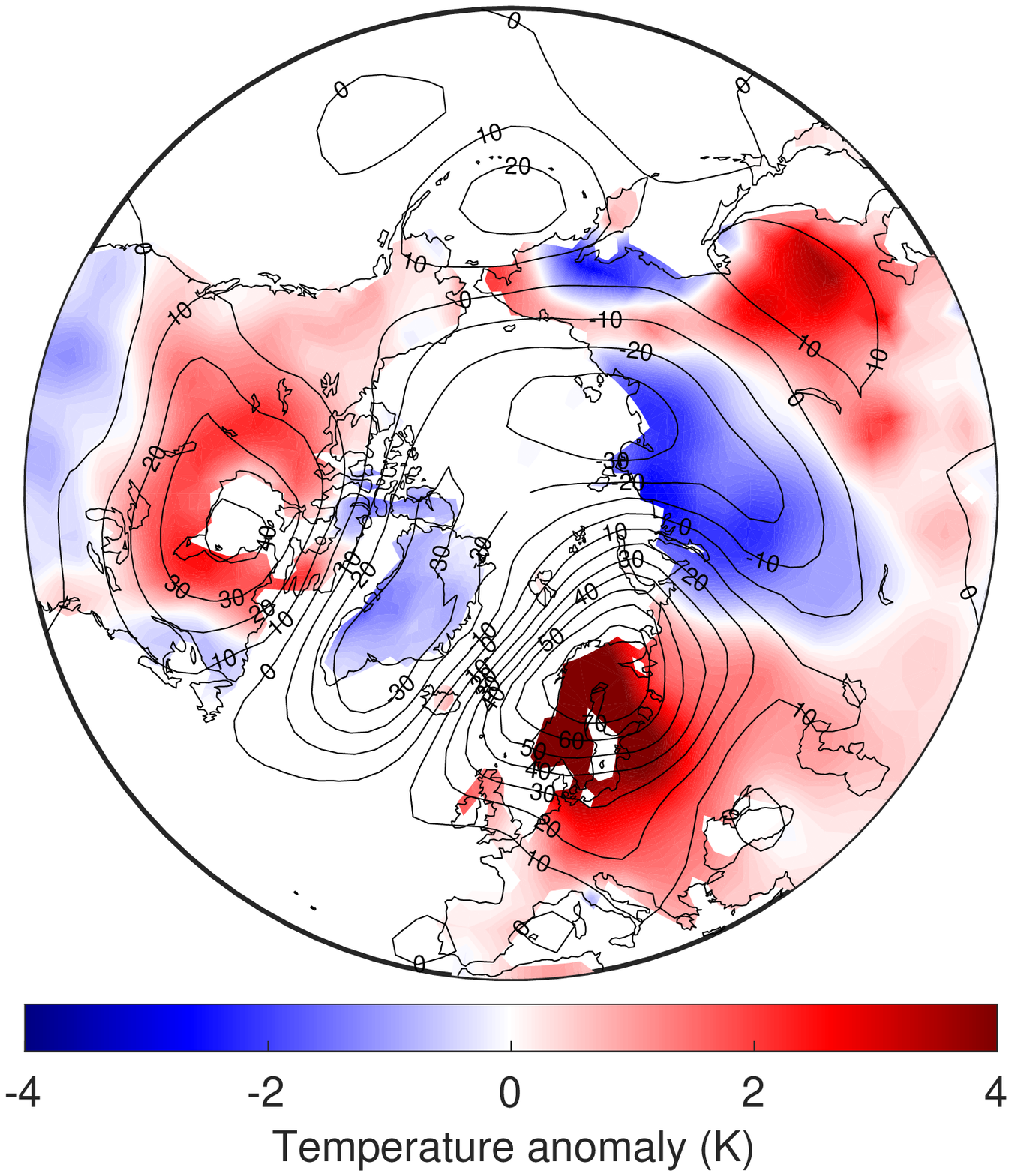}\\ 
(b)\includegraphics[width=0.45\textwidth]{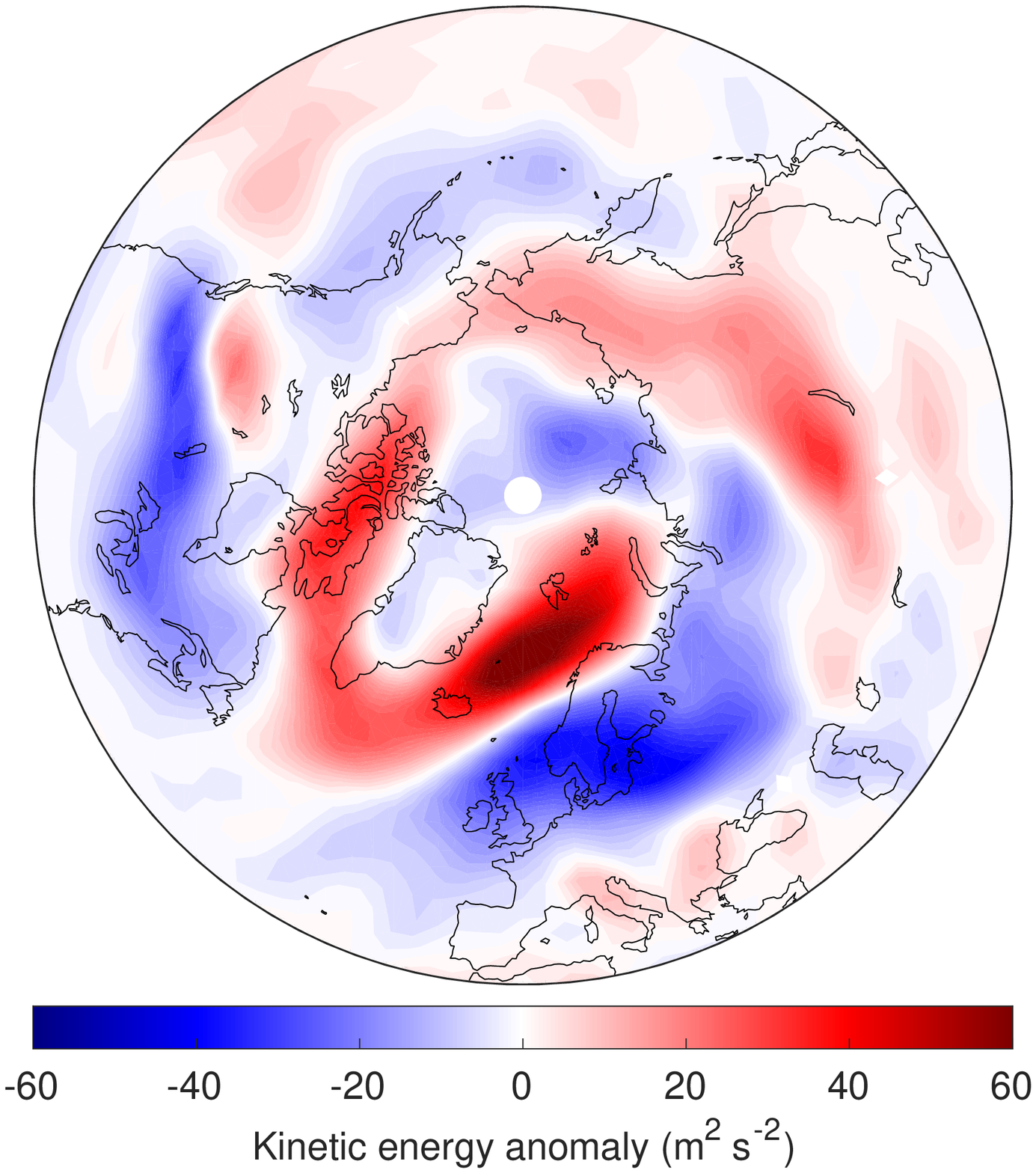}\\
\end{centering}
\caption{(a) Surface temperature anomaly (colors) and 500 hPa geopotential
height anomaly (contours) for the k50 experiment, conditional on the
occurrence of heat wave conditions. (b) Horizontal kinetic energy
anomaly at 500 hPa for the k50 experiment, conditional on the occurrence
of heat wave conditions.\label{fig:(a)-Surface-temperature}}
\end{figure}

Figure \ref{fig:(a)-Surface-temperature}a) shows the composite average
of surface temperature and 500 hPa geopotential height anomalies over
the Northern hemisphere above 35° latitude, conditional on the occurrence
of an European heat wave, estimated from the large deviation algorithm
with $k^{*}=50$. The heat wave pattern shows an extended warming
over Europe. The warm anomaly is larger over Scandinavia than over
the rest of Europe. The 500 hPa geopotential height field show a strong
anticyclonic anomaly right above the area experiencing the maximum
warming, as expected in heat wave conditions. Overall the pattern
over Europe is qualitatively very similar to the Scandinavian heat
wave cluster detected by \cite{Stefanon&al2012}. However, it appears
connected to a teleconnection pattern spanning the entire Northern
Hemisphere, with an apparent wavenumber 3 structure. The 500 hPa geopotential
height anomalies in the North Atlantic/European area are consistent
with a southward shift of the jet stream over the North Atlantic and
a northward shift of the jet stream over continental Europe, as shown
in \ref{fig:(a)-Surface-temperature} b), where we plot the composite
of the kinetic energy related to the horizontal motion, as anomalies
with respect to the control run.

Note that usual teleconnection patterns are computed typically through
empirical orthogonal functions analysis or similar techniques, and
thus describe pattern for typical fluctuations of the atmosphere.
Extreme event conditional statistics are instead related to very rare
states of the flow characterizing the extreme events. With $T$=90
days and $a$=2 K, the events that we have selected have return times
larger than 1000 years. Thanks to this method it is thus possible
to compute these maps with a sufficient degree of precision to be
able to robustly speak of teleconnection patterns for extreme events.

\section{Conclusions\label{sec:Conclusions}}

In this paper we have discussed techniques to compute large deviation
functions of climatic observables. Direct estimation of large deviation
functions in a complex chaotic system is a delicate procedure that
needs care in properly checking the convergence of both the large
deviation limit and of the statistical estimators \cite{Rohwer&al2015}.
A simplistic analysis relying on visual arguments for the collapse
of the scaled functions on some seemingly well behaved function can
lead to wrong estimates, as the functions will indeed converge, but
to wrong values. In particular, wrong estimates may lead to think
that the available data were enough to go beyond the Gaussian regime,
while as we have seen there may be cases in which a more refined analysis
shows that non Gaussian behaviors can appear as an artifact of not
having properly studied the convergence.

The rare event algorithm described in this paper \cite{del_moral_genealogical_2005,giardina_direct_2006,lecomte_numerical_2007}
can greatly help to compute large deviation rate functions for large
values of the anomalies that cannot be accessed with a direct approach.
This method was never used for systems of the complexity of a numerical
climate model, until in \cite{Ragone&al2018} we have used it to
study European heat waves. In this paper we have used it for the explicit
task of computing large deviation rate functions, highlighting its
connection to large deviation theory. The recipe presented in this
paper can be easily replicated for climate studies on different observables
and with different climate models.

We have observed that the large deviation limit is not of direct interest
for studying real heat waves. From a physical point of view, the problem
is that in the model Plasim with physical parameterizations, the autocorrelation
function of the European temperature has a slow decaying tail that
involves time scales of about $30$ days. The consequence of this
slow decorrelation is that the large time asymptotics of the large
deviation rate function is not attained before a few years. This makes
the use of the large deviation asymptotics irrelevant for this case
for time averages of the order of a few months. The large deviation
limit could however be of practical relevance for quantities with
faster decaying autocorrelation functions, for example precipitation,
or for spatial averages of surface temperature over different regions.

Moreover, algorithms initially dedicated to the computation of large
deviations are still efficient to compute the probability of extreme
heat waves. The results show in this paper and in \cite{Ragone&al2018}
refer to a perpetual summer setup, with no time dependent external
forcing acting on the system. In the applications envisioned by \cite{giardina_direct_2006}
and \cite{lecomte_numerical_2007}, the algorithm was not meant to
be applied in presence of a time dependent forcing on time scales
comparable with the duration of the events of interest, while in the
original formulation of \cite{del_moral_genealogical_2005} there
are not limitations in this sense. The daily cycle is shorter than
the resampling time, so that including it does not present any problem.
We are currently testing the performances of the algorithm in presence
of seasonal cycle, that will be the subject of future studies. Even
in the form presented in this paper however, the large deviation algorithm
could be of extreme interest for application to more theoretical studies
in which perpetual summer condition are a common setup.

\appendix

\section{Convergence of direct estimate of large deviation functions and statistical
errors\label{sec:Direct-estimate-Appendix}}

The choice of the size of the time block $\tau_{b}$ and the test
of the convergence to the large deviation limit requires computing
the autocorrelation time of the process, which sets a lower bound
to the values of the averaging time that it makes sense to consider.
Figure \ref{fig:a)-Probability-distribution}b) shows for the first
50 days the autocorrelation function $R(t)$ of the average European
surface temperature $T_{s}$ computed from a 1000 years long run.
We can see that to a first approximation the function is well described
by a double exponential, with a first decay time of about 4 days compatible
with the time scale of synoptic variability, followed by a slowly
decaying tail that at least in the first part seems to decay exponentially
on a time scale of one month. The longer time scales inducing the
slow decay of the autocorrelation function could be related to the
low frequency variability of the atmospheric dynamics, and/or to time
scales relate to the water vapor cycle in the atmosphere and the land
surface processes.

The integral autocorrelation time $\tau_{c}$ is defined as the integral
from time lag 0 to $+\infty$ of the autocorrelation function $R(t)=\mathbb{E}\left[(A(X(t))-\mu)(A(X(0))-\mu)\right]/\sigma^{2}$.
An equivalent expression for $\tau_{c}$ is \cite{Bouchet&al2015}

\begin{equation}
\tau_{c}=\frac{1}{2\sigma^{2}}\lim_{\tau_{b}\rightarrow+\infty}\frac{1}{\tau_{b}}\int_{0}^{\tau_{b}}\int_{0}^{\tau_{b}}\mathbb{E}\left[\left(A(X(t))-\mu\right)\left(A(X(s))-\mu\right)\right]dtds.\label{eq:autocorr_time_alt}
\end{equation}

\begin{figure}
\begin{centering}
\includegraphics[height=0.45\textwidth]{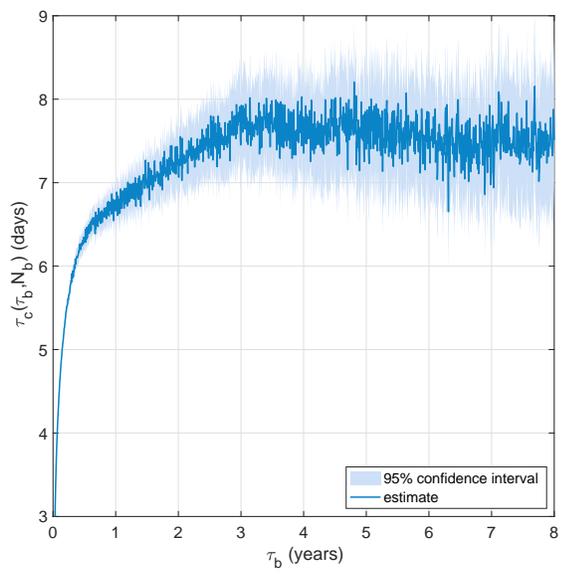}
\end{centering}
\caption{Convergence with $\tau_{b}$ of autocorrelation time.\label{fig:Convergence-with--1}}
\end{figure}

Equation \ref{eq:autocorr_time_alt} gives a better estimator of the
autocorrelation time than a simple time integration of the autocorrelation
function. In practice what we do is to divide the time series in $N_{b}$
blocks of length $\tau_{b}$, and then we compute the integrals in
(\ref{eq:autocorr_time_alt}) in each block and approximate the expectation
value as a sum of the $N_{b}$ blocks. Figure \ref{fig:Convergence-with--1}
shows the value of the estimate of the autocorrelation time as a function
of $\tau_{b}$. The shaded area represents the 95\% confidence interval
of the estimate computed as two standard deviations of the sample
of estimates over the $N_{b}$ blocks. We can see that the estimate
converges to a value of about 7.5 days, but that it is necessary to
use a very large value of $\tau_{b}$, of at least 3 years, in order
to reach convergence.

Once computed the autocorrelation time, the first step of the direct
estimate is to compute the generating function

\begin{equation}
\hat{G}(k,\tau_{b},N_{b})=\frac{1}{N_{b}}\sum_{j=1}^{N_{b}}e^{kA_{\tau_{b}}^{j}},\mathrm{with}\;A_{\tau_{b}}^{j}=\frac{1}{\tau_{b}}\intop_{(j-1)\tau_{b}}^{j\tau_{b}}A(X(t))dt\label{eq:GF_empirical-1}
\end{equation}
knowing that $\tau_{b}$ will have to be much larger than $\tau_{c}$.
When we deal with a discrete time series as the output of a numerical
model, where time is discretized in time steps of length $\Delta t$,
this means in practice computing

\begin{equation}
\hat{G}(k,\tau_{b},N_{b})=\frac{1}{N_{b}}\sum_{j=0}^{N_{b}-1}e^{k\sum_{n=jp+1}^{jp+p}A(X(n\Delta t))\Delta t},\label{eq:SCGF_empirical_simple}
\end{equation}
where $p=\tau_{b}/\Delta t$. Following \cite{Bouchet&al2015}, a
more sophisticated way that makes a better use of the available data
would be to compute

\begin{equation}
\hat{G}(k,\tau_{b},N_{b})=\frac{1}{2N_{b}}\sum_{j=0}^{2N_{b}-1}e^{k\sum_{n=jp/2+1}^{jp/2+p}A(X(n\Delta t))\Delta t}.\label{eq:SCGF_empirical_estimator}
\end{equation}
In (\ref{eq:SCGF_empirical_estimator}) the sample mean is computed
on $2N_{b}$ blocks overlapping by 50\%, as suggested by the Welch's
estimator of the power spectrum of a random process \cite{Welch1967}.
Using (\ref{eq:SCGF_empirical_estimator}) instead of (\ref{eq:SCGF_empirical_simple})
does not change the results of the estimate or the convergence region,
but gives smaller statistical errors where they can be computed. In
the following we keep the simpler notation (\ref{eq:GF_empirical-1})
for ease of presentation.

As discussed in the main text, in a practical application one is constrained
by the fixed length $T$ of the time-series, and the choice of $\tau_{b}$
and $N_{b}$ has to be considered carefully. The convergence of the
estimators has been studied by \cite{Rohwer&al2015}. In the case
of unbounded variables, obtaining a correct estimate is limited by
two problems: 1) the artificial linearization of the tails of the
functions due to the finite size of the sample and 2) the non-uniform
convergence for different values of $k$.

The linearization effect is an artifact in the estimate of $\hat{G}(k,\tau_{b},N_{b})$
for large values of $k$ which causes the estimate of $\hat{\lambda}(k,\tau_{b},N_{b})$
to become linear in $k$ for any value of $k$ whose module is large
enough. This is due to the fact that a sum of exponentials over a
finite sample, as the one involved in (\ref{eq:GF_empirical-1}),
is dominated for large $k$ by the largest value in the sample, so
that $\sum_{j=1}^{N_{b}}e^{kA_{\tau_{b}}^{j}}\approx e^{kA_{\tau_{b}}^{max}}$,
with $A_{\tau_{b}}^{max}=\max_{j}\{A_{\tau_{b}}^{j}\}$. Therefore,
for a given pair of $\tau_{b}$ and $N_{b}$, for positive $k$ there
is an upper critical value $k_{c}^{+}(\tau_{b},N_{b})>0$ for which
$\hat{\lambda}(k,\tau_{b},N_{b})\approx kA_{\tau_{b}}^{max}$ for
$k>k_{c}^{+}(\tau_{b},N_{b})$. Equivalently for negative $k$ there
is a lower critical value $k_{c}^{-}(\tau_{b},N_{b})<0$ for which
$\hat{\lambda}(k,\tau_{b},N_{b})\approx kA_{\tau_{b}}^{min}$ for
$k<k_{c}^{-}(\tau_{b},N_{b})$. If an observable is bounded, the linear
behavior is actually correct. For unbounded variables it is instead
an artifact of the finite size of the sample.

Scaling arguments can be provided to estimate $k_{c}^{+}(\tau_{b},N_{b})$
and $k_{c}^{-}(\tau_{b},N_{b})$, as discussed in details in \cite{Rohwer&al2015}.
However, the actual values depend on the underlying probability distribution
of the process, and in complex applications they have to be estimated
case by case by empirical analysis. A simple way to proceed is to
compute the relative contribution of the largest value to the sample
mean

\begin{equation}
r(k,\tau_{b},N_{b})=\frac{e^{kA_{\tau_{b}}^{max}}}{\sum_{j=1}^{N_{b}}e^{kA_{\tau_{b}}^{j}}}.\label{eq:indicator_linearization-1}
\end{equation}

\begin{figure}
\begin{centering}
a)\includegraphics[height=0.45\textwidth]{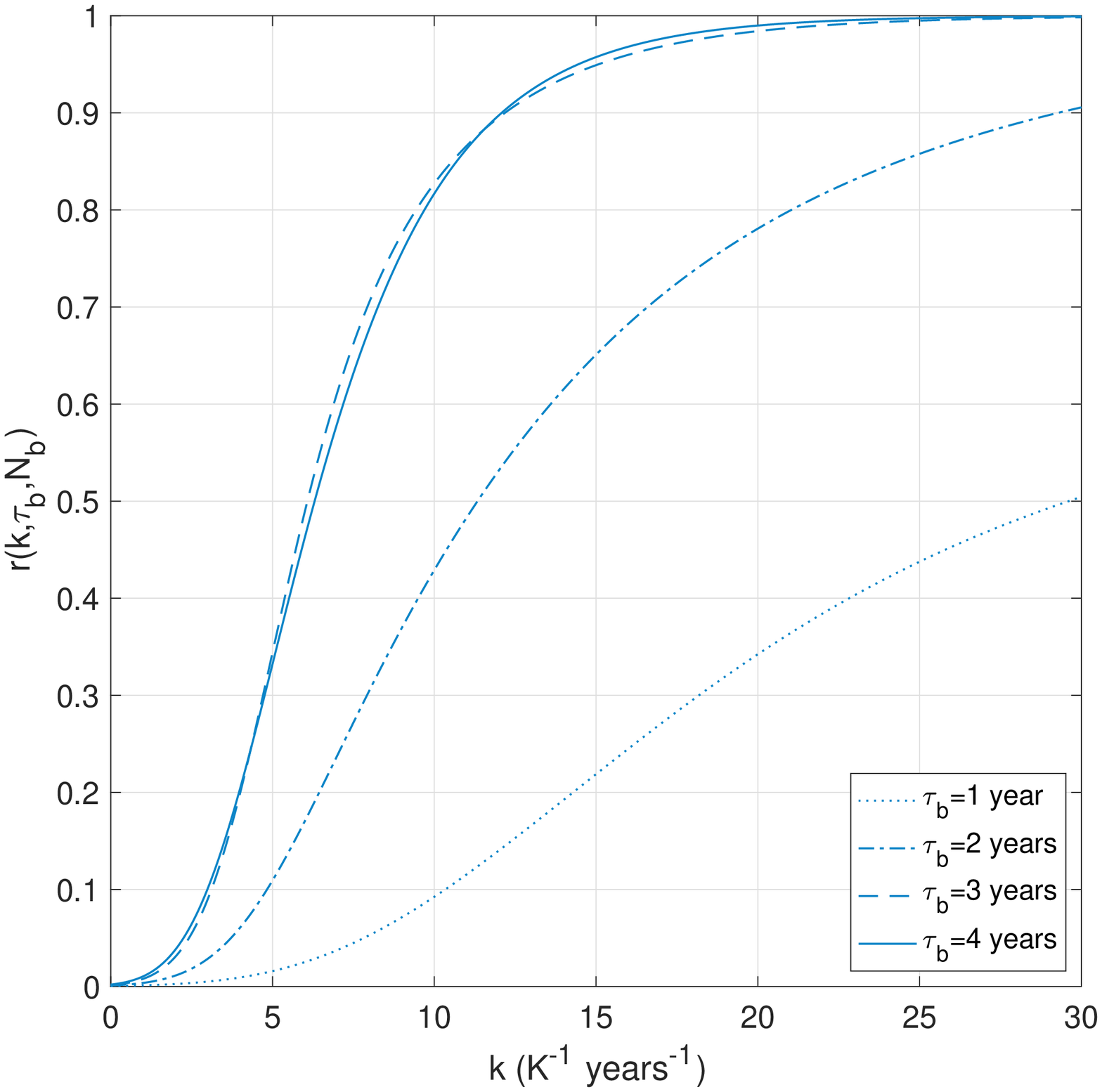}\\ 
b)\includegraphics[height=0.45\textwidth]{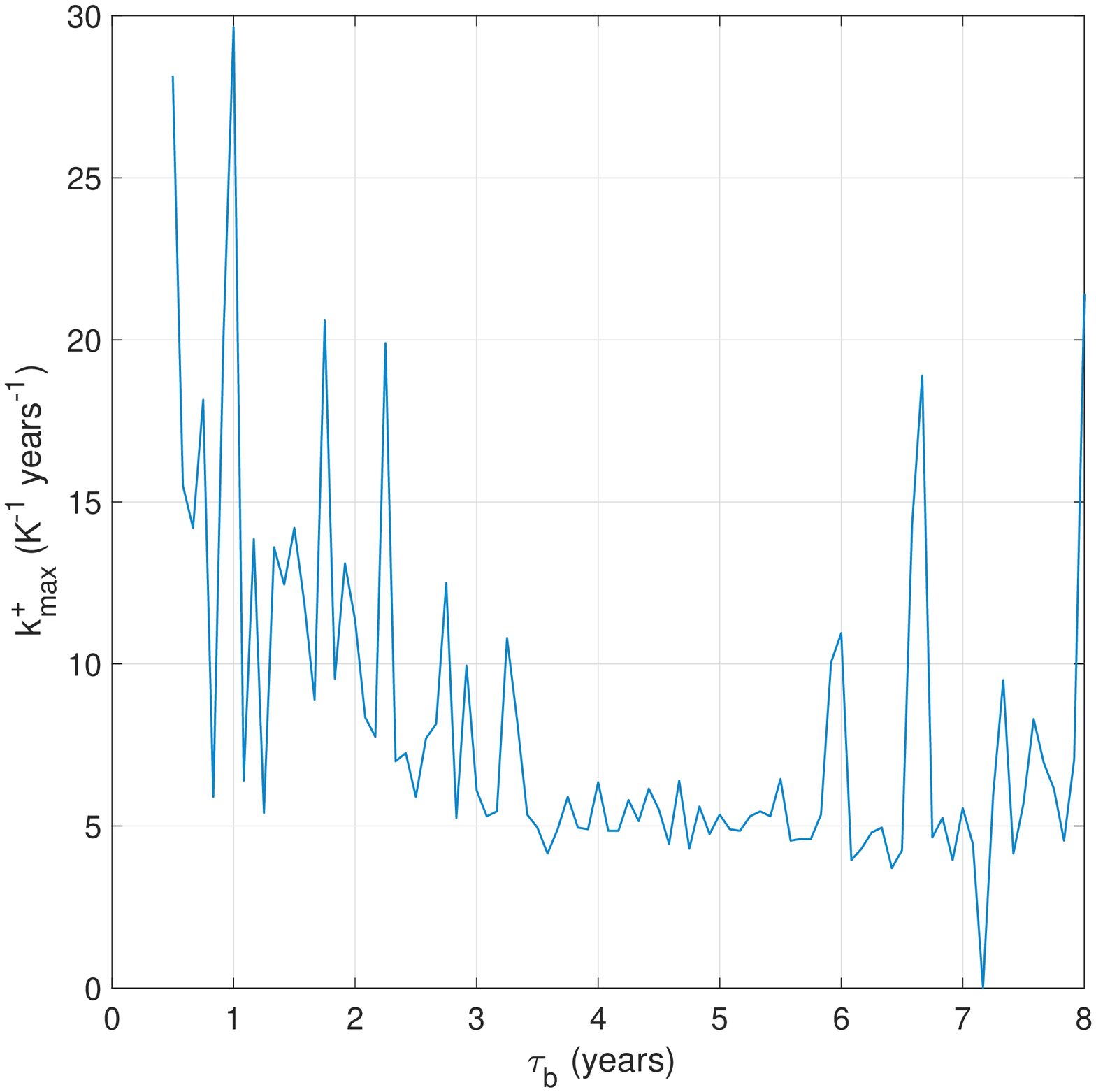}\\
\end{centering}
\caption{(a) Contribution of the largest value in the sample to the estimate
of the generating function as a function of $k$ for different values
of $\tau_{b}$. (b) Estimate of upper convergence limit as a function
of $\tau_{b}$, taking as threshold a 50\% contribution from the largest
value in the sample.\label{fig:(a)-Contribution-of}}
\end{figure}

By fixing an arbitrary upper threshold for $r(k,\tau_{b},N_{b})$,
one finds an estimate for the value of $k_{c}^{+}(\tau_{b},N_{b})$
(and an equivalent procedure gives a value for $k_{c}^{-}(\tau_{b},N_{b})$
). Figure \ref{fig:(a)-Contribution-of}a) shows $r(k,\tau_{b},N_{b})$
as a function of $k$ for different values of $\tau_{b}$ for which
there is actual convergence to the large deviation limit. Figure \ref{fig:(a)-Contribution-of}b)
shows the estimate of $k_{c}^{+}(\tau_{b},N_{b})$ as a function of
$\tau_{b}$, obtained taking a threshold of 50\% for $r(k,\tau_{b},N_{b})$.
We can see that there is a large difference in $k_{c}^{+}(\tau_{b},N_{b})$
if taking a value of $\tau_{b}$ of about 1 year or 3-4 years. However,
the estimate for lower values of $\tau_{b}$ is extremely unstable,
showing that if proper convergence in time is not reached, also the
convergence of the statistical estimator itself is not well behaved.
For $\tau_{b}$ larger than 3 years the estimate of $k_{c}^{+}(\tau_{b},N_{b})$
stabilizes around a value of 5 $K^{-1}years^{-1}$. We have therefore
taken $\tau_{b}=3$ years and $k_{c}^{+}(\tau_{b},N_{b})=5\:K^{-1}years^{-1}.$
A similar analysis gives $k_{c}^{-}(\tau_{b},N_{b})=-2.5\:K^{-1}years^{-1}$.

Once identified the convergence region of, one can compute statistical
errors in half of it, following \cite{Rohwer&al2015}. The error
on the generating function can be naturally estimated as

\begin{equation}
\mathsf{\mathfrak{\textrm{err}}}[\hat{G}(k,\tau_{b},N_{b})]=\sqrt{\mathrm{var}(\hat{G}(k,\tau_{b},N_{b}))/N_{b}},
\end{equation}
where $\mathrm{var}(\hat{G}(k,\tau_{b},N_{b}))$ is the empirical
variance associated with the sample mean replacing the expectation
value. An estimate of the associated error on $\hat{\lambda}(k,\tau_{b},N_{b})$
can be computed by taking a Taylor expansion of the estimator \cite{Pohorille&al2010,Rohwer&al2015} 

\begin{equation}
\mathsf{\mathfrak{\textrm{err}}}[\hat{\lambda}(k,\tau_{b},N_{b})]=\frac{\mathsf{\mathfrak{\textrm{err}}}[\hat{G}(k,\tau_{b},N_{b})]}{\hat{G}(k,\tau_{b},N_{b})}.
\end{equation}
The statistical error on $\hat{a}(k,\tau_{b},N_{b})$ can be estimated
by

\begin{equation}
\mathsf{\mathfrak{\textrm{err}}}[\hat{a}(k,\tau_{b},N_{b})]=\sqrt{\frac{\mathsf{\mathfrak{\textrm{err}}}[\hat{H}(k,\tau_{b},N_{b})]^{2}}{\left(\hat{G}(k,\tau_{b},N_{b})\right)^{2}}+\frac{\left(\hat{H}(k,\tau_{b},N_{b})\right)^{2}\mathsf{\mathfrak{\textrm{err}}}[\hat{G}(k,\tau_{b},N_{b})]^{2}}{\left(\hat{G}(k,\tau_{b},N_{b})\right)^{4}}},\label{eq:err_a-1}
\end{equation}
where $\hat{H}(k,\tau_{b},N_{b})=\sum_{j=1}^{N_{b}}A_{\tau_{b}}^{j}e^{kA_{\tau_{b}}^{j}}$
and $\mathsf{\mathfrak{\textrm{err}}}[\hat{H}(k,\tau_{b},N_{b})]$
is computed as $\mathsf{\mathfrak{\textrm{err}}}[\hat{G}(k,\tau_{b},N_{b})]$.
This formula is obtained assuming that $\hat{H}(k,\tau_{b},N_{b})$
and $\hat{G}(k,\tau_{b},N_{b})$ are independent \cite{Rohwer&al2015}.
The error on $\hat{I}(\hat{a}(k,\tau_{b},N_{b}),\tau_{b},N_{b})$
can then be estimated as

\begin{equation}
\mathsf{\mathfrak{\textrm{err}}}[\hat{I}(\hat{a}(k,\tau_{b},N_{b}),\tau_{b},N_{b})]=\sqrt{k\mathsf{\mathfrak{^{2}\textrm{err}}}[\hat{a}(k,\tau_{b},N_{b})+\mathsf{\mathfrak{\textrm{err}}}[\hat{\lambda}(k,\tau_{b},N_{b})}.
\end{equation}
again assuming independence between $\hat{a}(k,\tau_{b},N_{b})$ and
$\hat{\lambda}(k,\tau_{b},N_{b})$.

\section*{Acknowledgements}
The research leading to these results has received funding from the
European Research Council under the European Union\textquoteright s
seventh Framework Programme (FP7/2007-2013 Grant Agreement No. 616811). The simulations have been performed on the machines of the Pôle Scientifique de Modélisation Numérique (PSMN) and of the Centre Informatique National de l'Enseignement Supérieur (CINES).

\section*{Conflict of interest}
The authors declare that they have no conflict of interest.



\begin{thebibliography}{10}
\providecommand{\url}[1]{{#1}}
\providecommand{\urlprefix}{URL }
\expandafter\ifx\csname urlstyle\endcsname\relax
  \providecommand{\doi}[1]{DOI~\discretionary{}{}{}#1}\else
  \providecommand{\doi}{DOI~\discretionary{}{}{}\begingroup
  \urlstyle{rm}\Url}\fi

\bibitem{aghakouchak_extremes_2012}
AghaKouchak, A.: Extremes in a changing climate detection, analysis and
  uncertainty.
\newblock Springer, Dordrecht; New York (2012)

\bibitem{Ailliot&al2015}
Ailliot, P., Allard, D., Monbet, V., Naveau, P.: Stochastic weather generators:
  an overview of weather type models.
\newblock Journal de la Societe Francaise de Statistique \textbf{156}(1),
  101--113 (2015)

\bibitem{Bouchet&al2015}
Bouchet, F., Marston, J.B., Tangarife, T.: Fluctuations and large deviations of
  reynolds stresses in zonal jet dynamics.
\newblock Physics of Fluids \textbf{30}(1), 015110 (2018).
\newblock \doi{10.1063/1.4990509}

\bibitem{bucklew_introduction_2004}
Bucklew, J.A.: An introduction to rare event simulation.
\newblock Springer, New York (2004)

\bibitem{Coles2001}
Coles, S.: An Introduction to Statistical Modeling of Extreme Values.
\newblock Springer-Verlag, New York (2001)

\bibitem{delmoral_feynmankac_2004}
Del~Moral, P.: Feynman-Kac Formulae Genealogical and Interacting Particle
  Systems with Applications.
\newblock Springer New York (2004)

\bibitem{del_moral_genealogical_2005}
Del~Moral, P., Garnier, J.: Genealogical particle analysis of rare events.
\newblock The Annals of Applied Probability \textbf{15}(4), 2496--2534 (2005).
\newblock \doi{10.1214/105051605000000566}

\bibitem{dembo2001large}
Dembo, A., Zeitouni, O.: Large Deviations and Applications, in Handbook of
  stochastic analysis and application.
\newblock CRC Press (2001)

\bibitem{donsker1975asymptotic}
Donsker, M.D., Varadhan, S.S.: Asymptotic evaluation of certain markov process
  expectations for large time, i.
\newblock Communications on Pure and Applied Mathematics \textbf{28}(1), 1--47
  (1975)

\bibitem{ellis2007entropy}
Ellis, R.S.: Entropy, large deviations, and statistical mechanics.
\newblock Springer (2007)

\bibitem{Fischer&al2007}
Fischer, E.M., Seneviratne, S.I., Vidale, P.L., Luthi, D., Schaer, C.: Soil
  moisture-atmosphere interactions during the 2003 european summer heat wave.
\newblock Journal of Climate \textbf{20}(20), 5081--5099 (2007).
\newblock \doi{10.1175/JCLI4288.1}

\bibitem{Fraedrich&al2005}
Fraedrich, K., Jansen, H., Luksch, U., Lunkeit, F.: The {P}lanet {S}imulator:
  towards a user friendly model.
\newblock Meteorol. Z. \textbf{14}, 299--304 (2005)

\bibitem{Galfi&al2019}
Galfi, V.M., Lucarini, V., Wouters, J.: A large deviation theory-based analysis
  of heat waves and cold spells in a simplified model of the general
  circulation of the atmosphere.
\newblock Journal of Statistical Mechanics: Theory and Experiment
  \textbf{2019}(3), 033404 (2019).
\newblock \doi{10.1088/1742-5468/ab02e8}

\bibitem{Ghil&al2011}
Ghil, M., Yiou, P., Hallegatte, S., Malamud, B., Naveau, P., Soloviev, A.,
  Friederichs, P., Keilis-Borok, V., Kondrashov, D., Kossobokov, V., Mestre,
  O., Nicolis, C., Rust, H., Shebalin, P., Vrac, M., Witt, A., Zaliapin, I.:
  Extreme events: dynamics, statistics and prediction.
\newblock Nonlinear Processes in Geophysics \textbf{18}(3), 295--350 (2011).
\newblock \doi{10.5194/npg-18-295-2011}

\bibitem{giardina_simulating_2011}
Giardina, C., Kurchan, J., Lecomte, V., Tailleur, J.: Simulating rare events in
  dynamical processes.
\newblock Journal of Statistical Physics \textbf{145}(4), 787--811 (2011).
\newblock \doi{10.1007/s10955-011-0350-4}

\bibitem{giardina_direct_2006}
Giardina, C., Kurchan, J., Peliti, L.: Direct evaluation of large-deviation
  functions.
\newblock Physical Review Letters \textbf{96}(12), 120603 (2006).
\newblock \doi{10.1103/PhysRevLett.96.120603}

\bibitem{IPCC_managing_2012}
IPCC: Managing the risks of extreme events and disasters to advance climate
  change adaption: special report of the Intergovernmental Panel on Climate
  Change.
\newblock Cambridge University Press, New York, NY (2012)

\bibitem{IPCC_climate_2013}
IPCC: Climate Change 2013: The Physical Science Basis. Contribution of Working
  Group I to the Fifth Assessment Report of the Intergovernmental Panel on
  Climate Change.
\newblock Cambridge University Press, Cambridge, United Kingdom and New York,
  NY, USA (2013)

\bibitem{kahn_estimation_1951}
Kahn, H., Harris, T.E.: Estimation of particle transmission by random sampling.
\newblock National Bureau of Standards applied mathematics series \textbf{12},
  27--30 (1951)

\bibitem{kifer1990large}
Kifer, Y.: Large deviations in dynamical systems and stochastic processes.
\newblock Transactions of the American Mathematical Society \textbf{321}(2),
  505--524 (1990)

\bibitem{lecomte_numerical_2007}
Lecomte, V., Tailleur, J.: A numerical approach to large deviations in
  continuous time.
\newblock Journal of Statistical Mechanics: Theory and Experiment
  \textbf{2007}(03), P03004 (2007).
\newblock \doi{10.1088/1742-5468/2007/03/P03004}

\bibitem{Lestang&al2018}
Lestang, T., Ragone, F., Brehier, C.E., Herbert, C., Bouchet, F.: Computing
  return times or return periods with rare event algorithms.
\newblock Journal of Statistical Mechanics: Theory and Experiment
  \textbf{2018}(4), 043213 (2018).
\newblock \doi{10.1088/1742-5468/aab856}

\bibitem{Lorenz&al2010}
Lorenz, R., Jaeger, E.B., Seneviratne, S.I.: Persistence of heat waves and its
  link to soil moisture memory.
\newblock Geophysical Research Letters \textbf{37}(9), L09703 (2010).
\newblock \doi{10.1029/2010GL042764}

\bibitem{Lucarini&al2016book}
Lucarini, V., Faranda, D., Freitas, A.C.G.M.M., Freitas, J.M.M., Holland, M.,
  Kuna, T., Nicol, M., Todd, M., Vaienti, S.: Extremes and Recurrence in
  Dynamical Systems.
\newblock John Wiley and Sons, Inc. (2016)

\bibitem{Pohorille&al2010}
Pohorille, A., Jarzynski, C., Chipot, C.: Good practices in free-energy
  calculations.
\newblock J. Phys. Chem. B \textbf{114}, 10235--10253 (2010)

\bibitem{Ragone&al2018}
Ragone, F., Wouters, J., Bouchet, F.: Computation of extreme heat waves in
  climate models using a large deviation algorithm.
\newblock Proceedings of the National Academy of Sciences \textbf{115}(1),
  24--29 (2018).
\newblock \doi{10.1073/pnas.1712645115}

\bibitem{Rohwer&al2015}
Rohwer, C.M., Angeletti, F., Touchette, H.: Convergence of large-deviation
  estimators.
\newblock Phys. Rev. E \textbf{92}, 052104 (2015).
\newblock \doi{10.1103/PhysRevE.92.052104}

\bibitem{rubino_rare_2009}
Rubino, G., Tuffin, B.: Rare event simulation using Monte Carlo methods.
\newblock Wiley, Chichester, U.K. (2009)

\bibitem{Stefanon&al2012}
Stefanon, M., D'Andrea, F., Drobinski, P.: Heatwave classification over
  {E}urope and the {M}editerranean region.
\newblock Environmental Research Letters \textbf{7}, 014023 (2012)

\bibitem{Touchette2009}
Touchette, H.: The large deviation approach to statistical mechanics.
\newblock Physics Reports \textbf{478}, 1--69 (2009)

\bibitem{Veneziano&al2009}
Veneziano, D., Langousis, A., Lepore, C.: New asymptotic and preasymptotic
  results on rainfall maxima from multifractal theory.
\newblock Water Resources Research \textbf{45}(11), W11421 (2009).
\newblock \doi{10.1029/2009WR008257}

\bibitem{Welch1967}
Welch, P.D.: The use of fast {F}ourier transform for the estimation of power
  spectra: a method based on time averaging over short, modified periodograms.
\newblock IEEE Transactions on audio and electroacoustics \textbf{15}(2),
  70--73 (1967)

\bibitem{Wilks1999}
Wilks, D., Wilby, R.: The weather generation game: a review of stochastic
  weather models.
\newblock Progress in Physical Geography \textbf{23}(3), 329--357 (1999)

\bibitem{young1990large}
Young, L.S.: Large deviations in dynamical systems.
\newblock Transactions of the American Mathematical Society \textbf{318}(2),
  525--543 (1990)

\end{thebibliography}


\newpage

\end{document}